\documentclass[traditabstract]{aa}
\usepackage{txfonts}
\usepackage{natbib}
\usepackage{graphicx}

\bibpunct{(}{)}{;}{a}{}{,} 

\def\gsim{\mathrel{\hbox{\rlap{\lower.55ex \hbox {$\sim$}} \kern-.3em \raise.4ex \hbox{$>$}}}}
\def\lsim{\mathrel{\hbox{\rlap{\lower.55ex \hbox {$\sim$}} \kern-.3em \raise.4ex \hbox{$<$}}}}

\begin{document}

\title{The extremely low-metallicity tail of the Sculptor dwarf spheroidal galaxy \thanks{Based on observations collected at the European Organisation for Astronomical Research in the Southern Hemisphere, Chile proposal 085.D-0141.}}

\author{Else Starkenburg \inst{1,2,3} 
\and Vanessa Hill \inst{4} 
\and Eline Tolstoy \inst{1}
\and Patrick Fran\c{c}ois \inst{5,6}
\and Mike J. Irwin \inst{7} 
\and Leon Boschman \inst{1} 
\and Kim A. Venn \inst{2}
\and Thomas J. L. de Boer \inst{1}
\and Bertrand Lemasle \inst{1}
\and Pascale Jablonka \inst{5,8}
\and Giuseppina Battaglia \inst{9}
\and Paul Groot \inst{10}
\and Lex Kaper \inst{11}}

\institute{Kapteyn Astronomical Institute, University of Groningen, P.O. Box 800, 9700 AV Groningen, the Netherlands 
\and Department of Physics and Astronomy, University of Victoria, 3800 Finnerty Road, Victoria, BC, V8P 1A1, Canada \\
email: else@uvic.ca
\and CIfAR Junior Fellow and CITA National Fellow
\and Laboratoire Lagrange, UMR7293, Université de Nice Sophia-Antipolis, CNRS, Observatoire de la Côte d'Azur, 06300 Nice, France
\and GEPI, Observatoire de Paris, CNRS, Universit\'e Paris Diderot, Place Jules Janssen, 92190 Meudon, France
\and UPJV, Universit\'e de Picardie Jules Verne, 33 Rue St Leu, F-80080 Amiens, France
\and Institute of Astronomy, University of Cambridge, Madingley Road, Cambridge CB03 0HA, UK
\and Laboratoire d’astrophysique, \'Ecole Polytechnique F\'ed\'erale de Lausanne (EPFL), Observatoire de Sauverny, 1290 Versoix,
Switzerland
\and INAF – Osservatorio Astronomico di Bologna, via Ranzani 1, 40127 Bologna, Italy
\and Department of Astrophysics, Radboud University Nijmegen, P.O. Box 9010, 6500 GL Nijmegen, The Netherlands
\and Sterrenkundig Instituut ``Anton Pannekoek'', University of Amsterdam, Science Park 904, P.O. Box 94249, 1090 GE
Amsterdam, The Netherlands
}

\date{Received /
Accepted}

\abstract{We present abundances for seven stars in the (extremely) low-metallicity tail of the Sculptor dwarf spheroidal galaxy, from spectra taken with X-shooter on the ESO VLT. Targets were selected from the Ca II triplet (CaT) survey of the Dwarf Abundances and Radial Velocities Team (DART) using the latest calibration. Of the seven extremely metal-poor candidates, five stars are confirmed to be extremely metal-poor (i.e., [Fe/H]$<-3$ dex), with [Fe/H]=-3.47 $\pm$ 0.07 for our most metal-poor star. All have [Fe/H]$\leq-2.5$ dex from the measurement of individual Fe lines. These values are in agreement with the CaT predictions to within error bars. None of the seven stars is found to be carbon-rich. We estimate a 2--13\% possibility of this being a pure chance effect, which could indicate a lower fraction of carbon-rich extremely metal-poor stars in Sculptor compared to the Milky Way halo. The [$\alpha$/Fe] ratios show a range from +0.5 to --0.5, a larger variation than seen in Galactic samples although typically consistent within 1--2$\sigma$. One star seems mildly iron-enhanced. Our program stars show no deviations from the Galactic abundance trends in chromium and the heavy elements barium and strontium. Sodium abundances are, however, below the Galactic values for several stars. Overall, we conclude that the CaT lines are a successful metallicity indicator down to the extremely metal-poor regime and that the extremely metal-poor stars in the Sculptor dwarf galaxy are chemically more similar to their Milky Way halo equivalents than the more metal-rich population of stars.}


\authorrunning{Else Starkenburg et al.}
\titlerunning{The extremely low-metallicity tail of the Sculptor dwarf spheroidal galaxy}

\maketitle

\section{Introduction}

The low-metallicity stars that still exist today must carry the imprint of only very few supernovae explosions. These represent our closest observational approach to the epoch of the first stars as no completely heavy-element free star has been found to date. Many studies have been dedicated to the detailed investigation of extremely metal-poor stars in the Milky Way environment \citep[e.g.][]{bess84,norr01,cayr04,hond04,beer05,freb05,cohe06,fran07,cohe08,lai08,boni09,norr12,yong12a,yong12b}. An intriguing result from the very large and homogeneous study of \citet{cayr04}, is that various elements of the Milky Way population of very metal-poor stars ([Fe/H]$\le-2$ dex) and extremely metal-poor stars ([Fe/H]$\le-3$ dex) show little dispersion, indicating a cosmic scatter as low as 0.05 dex. However, a small percentage of stars in this regime have been shown to be chemically peculiar. Generally the stars with peculiar abundances prove to be carbon-rich, but also (a few) carbon-normal stars show abundance anomalies \citep[e.g.,][]{cohe08, yong12a}. 

An interesting puzzle was posed by the lack of extremely metal-poor stars found in the classical satellites around the Milky Way halo. Although these galaxies have very old stellar populations and a low average metallicity, no extremely metal-poor candidates were found from initial Ca II triplet (CaT, around 8500\ \AA) surveys, in contrast to the relative numbers of extremely metal-poor stars discovered in the Galactic halo \citep{helm06}. Various extremely metal-poor candidates were however discovered and followed-up using different selection methods, mainly in the ultra-faint dwarf galaxies \citep[e.g.,][]{kirb08,koch08,geha09,kirb09,norr10a,norr10b,freb10b,simo11,fran12}. The question thus arose if the empirical CaT method, extrapolated at [Fe/H]$<$--2.2 due to a lack of globular cluster measurement to calibrate, was biased in the metal-poor regime. \citet{star10} showed that none of the existing CaT calibrations to determine the metallicity for red giant branch (RGB) stars were reliable in the metal-poor regime ([Fe/H]$<-2.5$ dex). A new calibration was derived using synthetic spectral modeling tied to observations valid to [Fe/H]=--4.  Applying this to the Dwarf Abundances and Radial Velocities Team (DART) CaT datasets of $\sim$2000 stars in the Sculptor, Fornax, Carina and Sextans dwarf spheroidal galaxies resulted in many new candidate extremely metal-poor stars.  This analysis also brought the distribution of metal-poor stars in these dwarf galaxies in closer agreement with that of the Milky Way halo \citep{star10}.

To date, only samples of typically a few stars with [Fe/H]$<$--3 have been followed-up with high-resolution instruments in the classical galaxies Sculptor \citep{freb10a,tafe10}, Fornax \citep{tafe10}, Sextans \citep{aoki09,tafe10}, Ursa Minor \citep{kirb12} and Draco \citep{shet01,fulb04,cohe09}. The current lowest [Fe/H] for a star within a dwarf galaxy is [Fe/H]=$-3.96\pm0.06$ dex for a Sculptor star \citep{tafe10}.  In addition to [Fe/H], abundances for other elements are also very interesting since they highlight the contrast in chemical evolution at the earliest times between the Galactic halo and its surrounding satellites. The emerging picture is that metal-poor stars in the dwarf galaxies often have similar chemical abundance patterns to equivalent Galactic halo stars, unlike stars at higher metallicities where, for example, quite different [$\alpha$/Fe] ratios are found. On the other hand, some of these studies show evidence for inhomogeneous mixing, resulting in significant chemical variations within these small systems at the lowest metallicities and/or highest ages.  There are also indications of different abundance ratios for heavy elements of stars in the lowest mass systems \citep{freb10b,tafe10,venn12}.  

In this work we present the results of a follow-up study of seven extremely metal-poor candidates in the Sculptor dwarf spheroidal (Scl dSph) galaxy using the X-shooter spectrograph on the ESO VLT. We describe the target selection and observations in Sections \ref{sec:target} and \ref{sec:Xshootobs}. In Section \ref{sec:stelparam} we derive our stellar parameters from photometry and in Section \ref{sec:abundderive} we describe our abundance derivation method. We subsequently derive abundances for many elements in three comparison halo stars and our Sculptor targets in Sections \ref{sec:cayrab} and \ref{sec:sclab}. Our conclusions are presented in Section \ref{sec:Xshootconc}.

\section{Target selection}\label{sec:target}

\begin{figure}
\includegraphics[width=\linewidth]{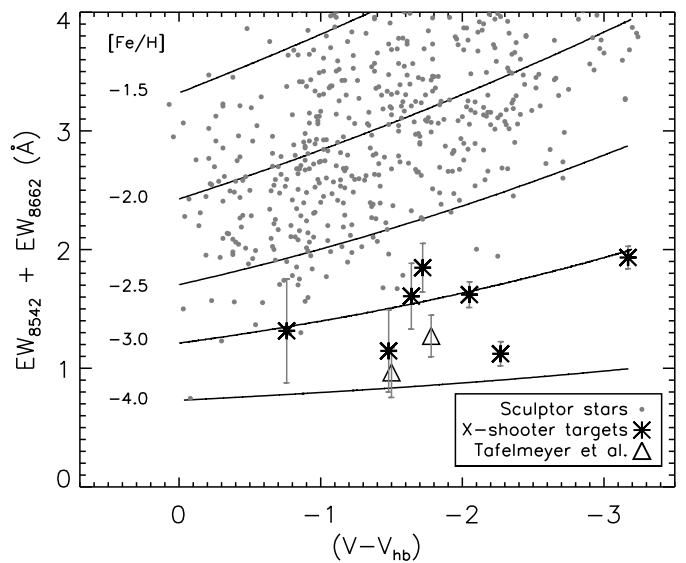}
\caption{The observed X-shooter targets in the Sculptor dwarf spheroidal (black asterisks) plotted as height above the horizontal branch versus equivalent width of the two strongest CaT lines. The error bars indicate the error in the equivalent width based on the observed S/N in the CaT observations. The complete sample of CaT candidates within the Sculptor dwarf galaxy are overplotted as small gray circles. The black open triangles are stars observed in previous high-resolution work by \citet{tafe10}. Overlaid as solid lines are the CaT calibration from \citet{star10}, where the corresponding [Fe/H] values are marked. \label{fig:targetsel}}
\end{figure}

\begin{figure}
\includegraphics[width=\linewidth]{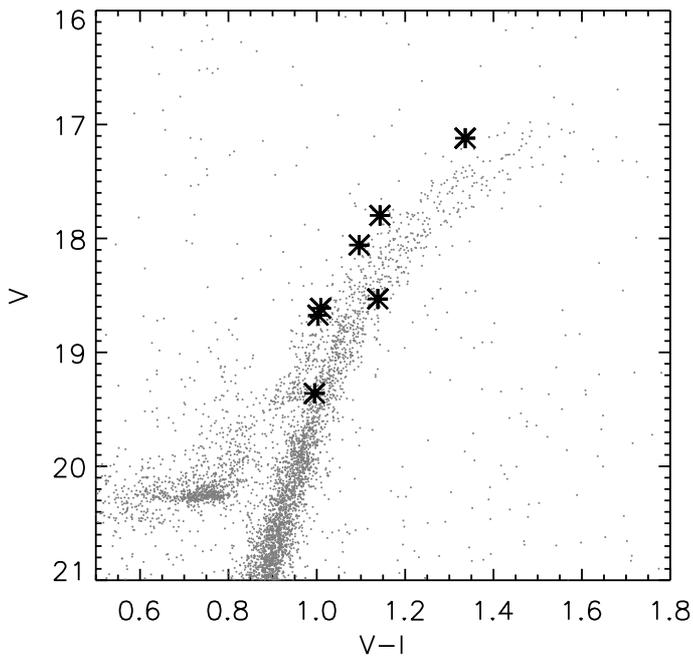}
\caption{A color-magnitude diagram of the inner regions of the Sculptor dwarf galaxy from \citet{debo11}, with the observed X-shooter targets overplotted as black asterisks. \label{fig:CMDtargetsel}}
\end{figure}

Targets were selected from the DART CaT survey of RGB stars in the Sculptor galaxy \citep{batt08a} which consists of CaT spectra taken with ESO FLAMES/GIRAFFE at a resolving power R$\sim$6500. Only radial velocity members of the Sculptor dwarf galaxy, i.e., stars with a radial velocity $<3\sigma$ from the systemic velocity of the Sculptor dwarf galaxy, were considered. Following \citet{batt08a} we use a systemic velocity $v_{hel,sys}=110.6$ km/s and a velocity dispersion $\sigma=10.1$ km/s. Due to the high latitude and distinct radial velocity of Sculptor from the Milky Way disk, the chance to accidentally select interlopers with this radial velocity criterion is very small \citep{batt12}. Figure \ref{fig:targetsel} shows the equivalent widths and absolute magnitude of the selected targets along with the calibration of \citet{star10} which deviates from earlier linear calibrations for very metal-poor stars. The targets were chosen to be extremely metal-poor candidates which were not already observed in other high-resolution studies. The targets are also selected over a range of absolute magnitudes (as can be seen in the color-magnitude diagram shown in Figure \ref{fig:CMDtargetsel}). This allows us to check the non-linear character of the relation of CaT equivalent width with luminosity and metallicity for the \citet{star10} calibration. However, limited observing time and weather constraints during the observing run meant that not all faint target candidates were observed. 

We used the star formation history derived by \citet{debo12} to compute the ages of the individual X-shooter targets from their positions in the CMD in Figure \ref{fig:CMDtargetsel}. All stars are consistent with both an old age ($\sim12$ Gyr) and a low metallicity ([Fe/H]$<-2.40$ dex), even though they show a spread on the RGB. Some targets overlap with the asymptotic giant branch, which is to be expected since this is where the extremely metal-poor giant branches also lie. If some of these stars are early asymptotic giant branch stars this would not significantly affect the abundance analysis presented here. 

\section{Observations}\label{sec:Xshootobs}
\subsection{Observations}
Observations were carried out as part of the GTO program for the X-shooter instrument on the ESO VLT \citep{vern11}. X-shooter is a long wavelength coverage (300--2500 nm) medium resolution spectrograph mounted at the UT2 Cassegrain focus. The instrument consists of 3 arms: UVB (covering the wavelength range 300--559.5 nm); VIS (559.5--1024 nm); and NIR (1024--2480 nm). One spectrum of an extremely metal-poor candidate star in Sculptor, Scl031\_11, was taken in November 2009. Spectra for six other extremely metal-poor candidates of the Sculptor dSph were taken in September 2010. In addition, three extremely metal-poor halo stars from the UVES high-resolution sample described in \citet{cayr04} have been observed to calibrate our results. We observed in slit mode with slit widths of 0.8$^{\prime \prime}$, 0.7$^{\prime \prime}$ and 0.9$^{\prime \prime}$ in the UVB, VIS and NIR arms respectively. In this work we focus on the spectra taken with the UVB and VIS arms, which have a resolving power of R$\sim$6200 in the UVB and R$\sim$11000 in the VIS with our settings. Although we could not measure individual weak lines because of the limited resolution, the large wavelength coverage enables us to measure many of the stronger lines available in the spectrum. The medium resolution and exceptional throughput additionally make the X-shooter instrument very suitable for the study of stars which are too faint for typical high-resolution instruments like UVES, as is the case for our faintest targets. 

\subsection{Data reduction}
The spectra were reduced using the X-shooter pipeline \citep{goldoni06, modi10} which performs bias and background subtraction, cosmic ray hit removal \citep{vandokkum01},  sky subtraction \citep{kelson03}, flat-fielding, order extraction and merging. Although the data have been acquired mostly in 1x1 nodding mode, the spectra were not reduced using  the nodding  pipeline recipes. Instead, each spectrum was reduced separately in slit mode with a manual localization of the source and the sky. This method allowed an optimal extraction of the spectra leading to an efficient cleaning of the remaining cosmic ray hits and also resulted in a noticeable improvement in the S/N ratio. The main drawback is that the sky subtraction is not as efficient as in nodding mode, as it leaves residual sky lines which could affect some stellar absorption lines located in the red part of the VIS spectrum. We therefore only use absorption lines which are not blended with sky lines in this analysis.

\subsection{Continuum normalization}
The targets are all very metal-poor stars and therefore the absorption lines in the spectrum are expected to be weak, resulting in a continuum level which is well-defined. On the other hand, since the equivalent widths measured for the lines will be relatively small, a robust continuum normalization is critical for accurate measurements. We applied two techniques for continuum normalization, the first using a high-order cubic spline to fit the continuum using the \textit{continuum} task in the Image Reduction and Analysis Facility (IRAF)\footnote{IRAF is written and supported by the IRAF programming group at the  National Optical Astronomy Observatories (NOAO) in Tucson, Arizona.}. Secondly the continuum was placed using an iterative k-sigma clipped non-linear filter \citep{batt08a}. We found that the differences between the two methods were minimal and for each star we selected the method that gave the most robust results for all the FeI lines across the whole wavelength region. 

Comparison of the overlapping region between 5500--5600 \AA\ in both the UVB and VIS arm shows that it is particularly difficult to apply a good continuum normalization in the region at the end of either arm, even if the S/N is relatively high. Therefore we do not use any weak lines in the overlap region nor the extreme wavelength ends of both arms because of very low S/N in the blue ($\lsim$3800 \AA) and the large number of sky lines in the red ($\gsim9000$ \AA).   

\section{Stellar parameters}\label{sec:stelparam}

The B, V  and I photometry for our stars are from a deep wide-field imaging project \citep[][see Figure \ref{fig:CMDtargetsel}]{debo11}. These data are complemented with infrared photometry from the VISTA survey commissioning where possible (see Table \ref{tab:photcol}). Table \ref{tab:temps} gives the temperatures as derived from the observed colors using a reddening correction of E(B-V) = 0.018, E(V-I) = 0.023, E(V-J) = 0.041 and E(V-K) = 0.050 in the direction of the Sculptor dwarf galaxy \citep{schl98} and the calibration of \citet{rami05} for their lowest metallicity bin (--4$<$[Fe/H]$<$--2.5). In general the relation from \citet{rami05} gives a slightly lower temperature for these stars than the calibration from \citet{alon99} for [Fe/H]=--3, which is used by \citet{cayr04}. The average difference between both temperature scales is 84 K from B$-$V colors. The largest dispersion among the temperatures derived from \citet{rami05} calibrations is 89 K, for Scl031\_11.

\begin{table*}
\centering
\center{\caption{Photometry for all Sculptor targets, no reddening correction applied.} \label{tab:photcol}}
\centering
\begin{tabular}{lcccccc}
\hline
\hline 
ID & V & B & I & J & K \\
\hline 
Scl002\_06 & 17.120 $\pm$ 0.005  &  18.265 $\pm$ 0.003  &  15.784 $\pm$ 0.005 & -- & --\\
Scl074\_02 & 18.058 $\pm$ 0.003  &  19.017 $\pm$ 0.003  &  16.962 $\pm$ 0.005 & -- & -- \\
Scl\_03\_170 & 18.674 $\pm$ 0.002  &  19.482 $\pm$ 0.008  &  17.671 $\pm$ 0.002 & 16.975 & 16.447 \\
Scl\_25\_031 & 18.613 $\pm$ 0.005  &  19.432 $\pm$ 0.005  &  17.604 $\pm$ 0.006 & -- & --\\
Scl051\_05 & 19.360 $\pm$ 0.003  &  20.140 $\pm$ 0.005  &  18.365 $\pm$ 0.004 & 17.694 & 17.014\\
Scl024\_01 & 18.531 $\pm$ 0.005  &  -- $\pm$ --         &  17.393 $\pm$ 0.018 & --  & --\\ 
Scl031\_11 & 17.798 $\pm$ 0.006  &  18.765 $\pm$ 0.003  &  16.655 $\pm$ 0.006 & 15.929 & 15.198\\
\hline
\end{tabular}
\end{table*}

\begin{table*}
\centering
\caption{Photometrically derived parameters for all the Sculptor targets.}
\label{tab:temps}
\centering
\begin{tabular}{l|ccccc|c|c}
\hline
\hline
  & \multicolumn{5}{c|}{T$_{\textnormal{eff}}$ (K)} & log(g) & v$_{mic}$ \\
ID   & B--V & V--I & V--J & V--K & av. & phot & (km/s)\\
\hline 
 Scl002\_06 &     4337   &   4348  &    -- & --   & 4343 & 0.69 & 2.3\\
 Scl074\_02 &     4547   &   4643  &    -- & --   & 4595 & 1.21 & 2.1 \\ 
 Scl\_03\_170 &   4739   &   4837  &  4846 & 4778 & 4800 & 1.56 & 1.9\\
 Scl\_25\_031 &   4723   &   4779  &   --  & --   & 4751 & 1.51 & 1.9\\
 Scl051\_05 &     4773   &   4813  &  4929 & 4784 & 4825 & 1.85 & 1.8\\
 Scl024\_01 &      --    &   4556  &    -- & --   & 4556 & 1.38 & 2.1\\
 Scl031\_11 &     4568   &   4647  &  4775 & 4616 & 4651 & 1.14 & 2.1\\
\hline
\end{tabular}
\end{table*}

The surface gravities for the program stars are also obtained from photometry and calculated using the standard relation given in Equation \ref{eq:logg}. In this equation we use the following solar values: $\textnormal{log}g_{\odot}=4.44$, T$_{\textnormal{\scriptsize{eff}},\odot}$=5790 K and $M_{bol,\odot}=4.72$. We further assume that the mass of all the RGB stars is 0.8 M$_{\odot}$. The absolute bolometric magnitude of the program stars is derived using the V-band magnitude, a distance modulus (m--M$_{0}$)=19.68 for the Sculptor dwarf galaxy \citep{piet08} and the bolometric correction calibration from \citet{alon99}. 

\begin{equation}
\label{eq:logg}
\textnormal{log}g_{*} = \textnormal{log}g_{\odot} + \textnormal{log}\frac{M_{*}}{M_{\odot}}+4\ \textnormal{log}\frac{T_{\textnormal{\scriptsize{eff}},*}}{T_{\textnormal{\scriptsize{eff}},\odot}} + 0.4\ (M_{bol,*}-M_{bol,\odot})
\end{equation}

We use the spectroscopically derived values for the microturbulence velocities of very and extremely metal-poor giant stars by \citet{cayr04} to estimate the values of the microturbulent velocities for our program stars. We do this by fitting linear relations to the dependence of the microturbulence velocity on the effective temperature and the surface gravity of the stars in \citet{cayr04} and use these relations to find the microturbulence velocities for our program stars listed in Table \ref{tab:temps}. These values match well with the trends of spectroscopically derived microturbulent velocities for the samples of stars studied in \citet{bark05} and \citet{tafe10}, but are generally higher than the spectroscopically derived values for more metal-rich stars of similar effective temperature and surface gravity from \citet{grat00}.  

\subsection{Radial velocities}
Radial velocities were measured using the \textit{fxcor} task in IRAF to cross-correlating the observed spectrum with a synthetic template spectrum with similar stellar parameters. We find very good agreement between the radial velocities measured from the CaT lines in previous low-resolution spectroscopic work \citep{batt08b} and our radial velocities also measured from the CaT region of the spectrum. The mean difference between the current and previous measurement is 2.8 km/s, and the largest difference measured is 5.9 km/s. All stars have radial velocities consistent with membership of the Sculptor dwarf galaxy. 

\section{Determination of abundances}\label{sec:abundderive}

\subsection{Line measurements}
Our linelist was created starting from the compilation of \citet{tafe10}, which in turn is based on \citet{cayr04} and \citet{shet03}. Naturally, because our work has much lower resolution, the weakest lines in this linelist could not be used. In addition we have added some lines from a comparison with synthetic spectra over the full wavelength range, which were checked not to be blended in an average low-metallicity spectrum. Most abundances have been derived from the equivalent widths that were measured by hand by fitting a Gaussian profile using \textit{splot} in the IRAF package. The full linelist and all equivalent widths for our sample are given in Table \ref{tab:linelist_targets} published in the online material. To reduce the contribution of noise within our final measurement, we only use lines which have an equivalent width larger than 25 m\AA, in accordance with the lower limit given by the Cayrel formula \citep{cayr88} for the measurement uncertainty of our worst S/N in the spectra. Lines which were marginally blended, such that two components could distinctly be seen, were measured using the deblending option in \textit{splot}. However, lines that deviated significantly from a Gaussian, either because of a blend or noise, were discarded. An exception has been made for the blended, but very strong, Sr lines, at 4077 \AA\ and 4215 \AA\ whose abundances could still be derived through a comparison with synthetic spectra. All the lines for which abundances are not derived using equivalent widths are marked in Table \ref{tab:linelist_targets} in the online material. For all lines stronger than 200 m\AA, and for the Mg triplet, Ba and Na lines, the line profile was further checked by comparison to a synthetic spectrum, and the abundance was updated if necessary. For barium synthesis we use the isotope ratios as given by \citet{mcwi98}.

\onltab{3}{
\begin{table*}
\caption{\label{tab:linelist_targets}Linelist and equivalent widths.}
\begin{tabular}{cccc|c c c c c c c}
\hline
\hline
$\lambda$ & EL  & $\chi_{ex}$& log(gf) & & &  EW $\pm \Delta$EW (m\AA) & & & \\
   & &        &     & Scl002\_06 & Scl074\_02 & Scl\_03\_170 & Scl\_25\_031 & Scl051\_05 & Scl024\_01 & Scl031\_11 \\
\hline
\hline
  5889.951 & Na1 & 0.00 & 0.12 & 132 $\pm$ 7\tablefootmark{2} & 134 $\pm$ 13\tablefootmark{2} & 163 $\pm$ 13\tablefootmark{2} & -- & 115 $\pm$ 15\tablefootmark{2} & 186 $\pm$ 16\tablefootmark{2} & 72 $\pm$ 16\tablefootmark{2} \\
  5895.924 & Na1 & 0.00 & --0.18 & 122 $\pm$ 7\tablefootmark{2} & 136 $\pm$ 13\tablefootmark{2} & -- & 159 $\pm$ 14\tablefootmark{2} & 104 $\pm$ 15\tablefootmark{2} & 153 $\pm$ 16\tablefootmark{2} & --\\
\hline
  4702.991 & Mg1  & 4.34 & --0.67 & 30 $\pm$ 8 & 45 $\pm$ 14 & 26 $\pm$ 8 & -- & -- & 60 $\pm$ 15\tablefootmark{2} & -- \\
  5172.684 & Mg1 & 2.71 & --0.40 & 218 $\pm$ 13\tablefootmark{2} & 166 $\pm$ 21\tablefootmark{1} & 184 $\pm$ 15\tablefootmark{1} & 208 $\pm$ 27\tablefootmark{2} & 189 $\pm$ 15\tablefootmark{2} & 162 $\pm$ 22\tablefootmark{2} & 102 $\pm$ 22\tablefootmark{1} \\
  5183.604 & Mg1 & 2.71 & --0.18 & 188 $\pm$ 9\tablefootmark{2} & 203 $\pm$ 13\tablefootmark{1} & 174 $\pm$ 12\tablefootmark{2} & 200 $\pm$ 16\tablefootmark{2} & 200 $\pm$ 14\tablefootmark{2} & 236 $\pm$ 13\tablefootmark{2} & 121 $\pm$ 13\tablefootmark{1} \\
  8806.756 & Mg1 & 4.35 & --0.14  & 83 $\pm$ 7& 85 $\pm$ 10 & 69 $\pm$ 10 & 94 $\pm$ 12 & -- & 132 $\pm$ 18 & 14 $\pm$ 7\\
\hline
  6102.723 & Ca1 & 1.88 & --0.79 & --  & 28 $\pm$ 11 & -- & -- & -- & -- & -- \\ 
  6122.217 & Ca1 & 1.89 & --0.32  & 57 $\pm$ 8 & 69 $\pm$ 13 & 38 $\pm$ 12 & 55 $\pm$ 12 & 31 $\pm$ 12 & 49 $\pm$ 18 & -- \\ 
  6162.173 & Ca1 & 1.90 & --0.09  & 67 $\pm$ 7 & 74 $\pm$ 12 & 31 $\pm$ 9 & 70 $\pm$ 13 & -- & 69 $\pm$ 16 & -- \\ 
  6439.075 & Ca1 & 2.53 & 0.39 & 34 $\pm$ 7 & 49 $\pm$ 12 & 25 $\pm$ 19 & 22 $\pm$ 7 & 55 $\pm$ 12 & -- & -- \\ 
\hline
  4981.731 & Ti1 & 0.85 & 0.50 & -- & -- & -- & -- & -- & 48 $\pm$ 15 & -- \\  
  4991.065 & Ti1 & 0.84 & 0.38 & -- & -- & -- & 45 $\pm$ 16 & -- & 52 $\pm$ 10 & -- \\ 
  4999.503 & Ti1 & 0.83 & 0.25 & -- & -- & -- & -- & -- & 36 $\pm$ 13 & -- \\ 
  5064.653 & Ti1 & 0.05 & --0.99 & -- & -- & 38 $\pm$ 9 & -- & -- & -- & -- \\ 
  5210.385 & Ti1 & 0.05 & --0.88 & -- & -- & --  & 47 $\pm$ 11 & -- & 34 $\pm$ 9 & -- \\ 
\hline
  4468.507 & Ti2 & 1.13 & --0.60 & 132 $\pm$ 11 & 76 $\pm$ 14\tablefootmark{2} & 133 $\pm$ 13 & 133 $\pm$ 22 & 84 $\pm$ 13 & 132 $\pm$ 15 & 72 $\pm$ 22 \\ 
  4501.270 & Ti2 & 1.12 & --0.77 & 107 $\pm$ 11 & 60 $\pm$ 13\tablefootmark{2} & 105 $\pm$ 16  & 116 $\pm$ 33 & 113 $\pm$ 21 & 116 $\pm$ 14 & -- \\ 
  4563.757 & Ti2 & 1.22 & --0.69 & 98 $\pm$ 9 & -- & 84 $\pm$ 16 & 124 $\pm$ 23 & 75 $\pm$ 18 & 165 $\pm$ 15 & 84 $\pm$ 23 \\ 
  5129.156 & Ti2 & 1.89 & --1.24 & -- & -- & -- & 46 $\pm$ 18 & -- & 86 $\pm$ 16\tablefootmark{2} & --\\ 
  5154.068 & Ti2 & 1.57 & --1.75 & -- & -- & 32 $\pm$ 11 & -- & -- & -- & -- \\ 
  5185.902 & Ti2 & 1.89 & --1.49 & 31 $\pm$ 7 & -- & -- & -- & -- & 52 $\pm$ 15 & 35 $\pm$ 13 \\
  5188.687 & Ti2 & 1.58 & --1.05 & 63 $\pm$ 8 & -- & 68 $\pm$ 10 & -- & 38 $\pm$ 10 & -- & -- \\ 
  5336.786 & Ti2 & 1.58 & --1.59 & -- & -- & -- & -- & 25 $\pm$ 9 & 58 $\pm$ 12 & 27 $\pm$ 16 \\ 
\hline
  5206.037 & Cr1 & 0.94 & 0.02 & 78 $\pm$ 8 & 73 $\pm$ 7 & -- & 50 $\pm$ 18  & -- & 119 $\pm$ 14 & -- \\ 
  5208.425 & Cr1 & 0.94 & 0.16 & 80 $\pm$ 7 & -- & 80 $\pm$ 11 & 138 $\pm$ 17  & 122 $\pm$ 13 & 100 $\pm$ 12 & -- \\ 
  5296.691 & Cr1 & 0.98 & --1.40 & -- & -- & -- & -- & -- & -- & -- \\ 
  5345.796 & Cr1 & 1.00 & --0.98 & -- & -- & --  & -- & 31 $\pm$ 11 & -- & -- \\
  5409.784 & Cr1 & 1.03 & --0.72 & -- & -- & -- & 44 $\pm$ 17 & -- & -- & -- \\
\hline
  4071.737  & Fe1  & 1.61 & --0.02 & 183 $\pm$ 12 & 143 $\pm$ 22 & 87 $\pm$ 21 & 129 $\pm$ 55 & 179 $\pm$ 25 & 136 $\pm$ 27 & 147 $\pm$ 36 \\    
  4132.058  & Fe1  & 1.61 & --0.68 & 173 $\pm$ 14 & 114 $\pm$ 16 & 109 $\pm$ 20 & 51 $\pm$ 19  & 132 $\pm$ 28 & -- & 116 $\pm$ 35 \\    
  4404.750  & Fe1  & 2.40 & 0.11 & 176 $\pm$ 9 & 140 $\pm$ 14 & 104 $\pm$ 13 & 107 $\pm$ 23 & 174 $\pm$ 17 & 147 $\pm$ 14 & 113 $\pm$ 26\\    
  4461.652  & Fe1  & 0.09 & --3.21 & 171 $\pm$ 11 & 134 $\pm$ 14 & 65 $\pm$ 15 & 134 $\pm$ 20 & 102 $\pm$ 15 & 114 $\pm$ 15& 92 $\pm$ 23\\    
  4476.019  & Fe1 & 2.85 & --0.82 & -- & 36 $\pm$ 14 & 47 $\pm$ 15 & --  & 72 $\pm$ 20 & 71 $\pm$ 13 & 27 $\pm$ 18 \\    
  4494.563  & Fe1  & 2.20 & --1.14 & 107 $\pm$ 11 & 67 $\pm$ 13 & -- & 36 $\pm$ 19 & 131 $\pm$ 18 & 69 $\pm$ 14 & -- \\    
  4918.993  & Fe1 & 2.87 & --0.34 &  52 $\pm$ 10 & 49 $\pm$ 12 & -- & 48 $\pm$ 12 & 84 $\pm$ 15 & -- & -- \\    
  4920.502  & Fe1 & 2.83 & 0.07 & 102 $\pm$ 9 & 88 $\pm$ 13 & -- & 117 $\pm$ 14 & 111 $\pm$ 14 & 131 $\pm$ 12 & --\\    
  4994.129  & Fe1 & 0.92 & --3.08 & 86 $\pm$ 10 & 72 $\pm$ 12 & -- & 30 $\pm$ 17 & 34 $\pm$ 10 & 58 $\pm$ 12 & --\\    
  5012.067  & Fe1 & 0.86 & --2.64 & 138 $\pm$ 9 & 81 $\pm$ 11 & 83 $\pm$ 13 & 91 $\pm$ 16 & -- & 103 $\pm$ 12 & --\\    
  5049.819  & Fe1 & 2.28 & --1.36 & 86 $\pm$ 10 & 70 $\pm$ 16 & 33 $\pm$ 9 & -- & 63 $\pm$ 13 & 62 $\pm$ 13 & --\\    
  5051.634  & Fe1 & 0.92 & --2.80 & 129 $\pm$ 10 & 93 $\pm$ 11 & 70 $\pm$ 10 & 96 $\pm$ 16 & 76 $\pm$ 15 & 110 $\pm$ 13 & 32 $\pm$ 12 \\    
  5083.338  & Fe1 & 0.96 & --2.96 & 70 $\pm$ 8 & -- & 60 $\pm$ 12 & 32 $\pm$ 20 & 67 $\pm$ 13 & 110 $\pm$ 13 & -- \\    
  5110.413  & Fe1 & 0.00 & --3.76 & 154 $\pm$ 8 & 114 $\pm$ 12 & 73 $\pm$ 13 & -- & 107 $\pm$ 14 & 116 $\pm$ 12 & 63 $\pm$ 14 \\    
  5127.358  & Fe1 & 0.92 & --3.31 & 87 $\pm$ 12 & 58 $\pm$ 13 & -- & 33 $\pm$ 15 & 39 $\pm$ 12 & 79 $\pm$ 14 & --\\    
  5194.941  & Fe1 & 1.56 & --2.09 & -- & -- & 33 $\pm$ 9 & 79 $\pm$ 17 & 88 $\pm$ 15 & -- & --\\    
  5232.939  & Fe1 & 2.94 & --0.06 & 61 $\pm$ 8 & 65 $\pm$ 10 & 53 $\pm$ 10 & 73 $\pm$ 17 & 109 $\pm$ 13 & 80 $\pm$ 15 & 31 $\pm$ 10\\    
  5324.178  & Fe1 & 3.21 & --0.10 & 65 $\pm$ 8 & -- & -- & 66 $\pm$ 15 & 85 $\pm$ 15 & 71 $\pm$ 11 & 29 $\pm$ 10 \\    
  5371.489  & Fe1 & 0.96 & --1.65 & 180 $\pm$ 8 & 158 $\pm$ 9 & 121 $\pm$ 10 & 109 $\pm$ 14 & 163 $\pm$ 14 & 159 $\pm$ 12 & 118 $\pm$ 12 \\    
  5397.127  & Fe1 & 0.92 & --3.31 & 156 $\pm$ 7 & 139 $\pm$ 10 & 132 $\pm$ 11 & -- & 127 $\pm$ 12 & 148 $\pm$ 11 & 105 $\pm$ 12 \\    
  5405.774  & Fe1 & 0.99 & --1.84 & 154 $\pm$ 7 & 115 $\pm$ 9 & 90 $\pm$ 12 & 106 $\pm$ 13 & 146 $\pm$ 15 & 137 $\pm$ 11 & 106 $\pm$ 15 \\    
  5429.696  & Fe1 & 0.96 & --1.99 & 162 $\pm$ 8 & 144 $\pm$ 9 & 118 $\pm$ 12 & 120 $\pm$ 13 & -- & 140 $\pm$ 11 & 110 $\pm$ 13 \\   
  5434.523  & Fe1 & 1.01 & --2.12 & 160 $\pm$ 7 & 117 $\pm$ 8 & 113 $\pm$ 11 & -- & -- & 116 $\pm$ 10 & 93 $\pm$ 15 \\    
  5446.916  & Fe1 & 0.99 & --1.91 & 178 $\pm$ 7 & 154 $\pm$ 9 & 108 $\pm$ 10 & 101 $\pm$ 15 & -- & 169 $\pm$ 12 & 110 $\pm$ 13\\    
  5455.609  & Fe1 & 1.01 & --2.09 & 230 $\pm$ 8 & 153 $\pm$ 11 & 135 $\pm$ 11 & 136 $\pm$ 14 & 143 $\pm$ 12 & 151 $\pm$ 13 & 105 $\pm$ 14 \\      
  5615.644 & Fe1 & 3.33 & 0.05  & 52 $\pm$ 7 & -- & 70 $\pm$ 14 & 31 $\pm$ 11  & 87 $\pm$ 17 & 85 $\pm$ 16 & 41 $\pm$ 12 \\    
  6191.558  & Fe1 & 2.43 & --1.42 & -- & -- & 37 $\pm$ 13 & -- & 106 $\pm$ 18 & 73 $\pm$ 26 & --\\    
  6230.722  & Fe1 & 2.56 & --1.28 & 43 $\pm$ 8 & 26 $\pm$ 10 & -- & -- & 82 $\pm$ 18 & 39 $\pm$ 17 & --\\    
  6421.349  & Fe1 & 2.28 & --2.03 & -- & -- & -- & -- & 73 $\pm$ 18 & 32 $\pm$ 13 & --\\    
  6430.844  & Fe1 & 2.18 & --1.49 & 43 $\pm$ 8 & -- & -- & -- & -- & -- & 40 $\pm$ 16 \\    
  6494.980  & Fe1 & 2.40 & --1.27 & 60 $\pm$ 8 & 39 $\pm$ 10 & 46 $\pm$ 13 & 46 $\pm$ 12 & 82 $\pm$ 16 & 68 $\pm$ 20 & --\\    
  8327.052  & Fe1 & 2.20 & --1.53 & 111  $\pm$ 8 & -- & 103 $\pm$ 11 & -- & -- & 97 $\pm$ 18 & --\\    
  8387.770  & Fe1 & 2.18 & --1.49 & 68 $\pm$ 7 & 49 $\pm$ 9 & -- & 41 $\pm$ 8 & 71 $\pm$ 18 & 76 $\pm$ 18 & 37 $\pm$ 15\\    
  8468.404  & Fe1 & 2.22 & --2.07 & 42 $\pm$ 7 &  -- & -- & 35 $\pm$ 11 & -- & 34 $\pm$ 21 & --\\    
  8688.621  & Fe1 & 2.18 & --1.21 & 139 $\pm$ 7 & 105 $\pm$ 10 & 110 $\pm$ 12 & 71 $\pm$ 11 & 92 $\pm$ 18 & 125 $\pm$ 20 & --\\    
  8824.216  & Fe1 & 2.20 & --1.54 & -- & -- & -- & -- & -- & 50 $\pm$ 14 & --\\  
\hline
\multicolumn{9}{c}{Continued on next page}\\
\hline
\hline
\end{tabular}
\end{table*}
}

\onltab{3}{
\begin{table*}
\caption{continued.}
\begin{tabular}{cccc|c c c c c c c}
\hline
\hline
$\lambda$ & EL  & $\chi_{ex}$& log(gf) & & & &  EW $\pm \Delta$EW (m\AA) & & & \\
   & &        &     & Scl002\_06 & Scl074\_02 & Scl\_03\_170 & Scl\_25\_031 & Scl051\_05 & Scl024\_01 & Scl031\_11 \\
\hline
\hline
  4583.837 & Fe2  & 2.81 & --1.87 & 69 $\pm$ 10 & -- & -- & -- & 103 $\pm$ 17 & -- & -- \\  
  4923.927 & Fe2  & 2.89 & --1.50 & 111 $\pm$ 10 & 75 $\pm$ 12 & 76 $\pm$ 10 & 89 $\pm$ 20   & 75 $\pm$ 14 & 105 $\pm$ 12 & 70 $\pm$ 12 \\  
  5018.440 & Fe2  & 2.89 & --1.35 & 104 $\pm$ 9 & 90 $\pm$ 13 & 126 $\pm$ 11 & -- & 94 $\pm$ 14 & 96 $\pm$ 12 & 65 $\pm$ 14 \\  
  5276.002 & Fe2  & 3.20 & --2.21 & -- & 22 $\pm$ 11 & -- & -- & 56 $\pm$ 11 & 52 $\pm$ 11 & 27 $\pm$ 10\\  
\hline
  5476.900  & Ni1 & 1.83 & --0.89 & 61 $\pm$ 7 & -- & -- & -- & -- & 77 $\pm$ 12 & -- \\  
\hline
  4077.709 & Sr2 & & & \multicolumn{7}{c}{Blended line - abundance derived through comparison with synthetic spectra}\\
  4215.519 & Sr2  & & & \multicolumn{7}{c}{Blended line - abundance derived through comparison with synthetic spectra}\\
\hline
  4554.029 & Ba2 & 0.00 & 0.17 & -- & -- & 78 $\pm$ 14\tablefootmark{1} & -- & -- & --  & -- \\ 
  4934.076 & Ba2 & 0.00 & --0.15 & -- & -- & 67 $\pm$ 11\tablefootmark{1}  & -- & -- & 119 $\pm$ 12\tablefootmark{1} & -- \\ 
\hline
\hline
\end{tabular}
\tablefoot{\tablefoottext{1}{The derived abundance for this line is checked by direct comparison of the line shape to synthetic spectra.} \\ 
\tablefoottext{2}{Abundance is updated after comparison with synthetic spectra.}}
\end{table*}
}

\begin{table*}
\centering
\caption{Positions, CaT predictions for [Fe/H] and the S/N in various wavelengths of the spectra.\label{tab:targets}} 
\begin{tabular}{ccccc|ccccccc}
\hline
\hline
ID Sculptor star & $\alpha$ & $\delta$ & ellrad  & [Fe/H]$_{\textnormal{\scriptsize{CaT}}}$ & \multicolumn{7}{c}{S/N in wavelength region around (\AA)}  \\
   & (h m s J2000)  &  (d m s J2000) &  (deg)  & (dex)          & 4000 & 4600 & 5000 & 5300 & 6000 & 7000 & 8700\\
\hline 
Scl002\_06  & 01 01 26.74 &  --33 02 59.8 & 1.03 & --3.03 $\pm$ 0.09 & 39 & 65 & 71 & 85 & 71 & 89 & 95\\
Scl074\_02  & 00 57 34.84 &  --33 39 45.6 & 0.54 & --3.02 $\pm$ 0.09 & 26 & 50 & 50 & 68 & 41 & 53 & 58\\
Scl\_03\_170 & 01 01 47.46 &  --33 47 27.8 & 0.35 & --3.40 $\pm$ 0.68 & 23 & 54 & 56 & 57 & 38 & 43 & 49\\
Scl\_25\_031 & 00 59 41.05 &  --32 50 04.2 & 1.27 & --2.94 $\pm$ 0.26 & 13 & 26 & 35 & 40 & 26 & 31 & 27\\
Scl051\_05  & 00 59 25.96 &  --33 25 26.3 & 0.42 & --3.04 $\pm$ 0.70 & 20 & 34 & 42 & 44 & 31 & 42 & 44\\
Scl024\_01  & 01 02 41.23 &  --33 18 30.9 & 0.84 & --2.77 $\pm$ 0.26 & 21 & 41 & 46 & 54 & 31 & 52 & 48\\
Scl031\_11  & 00 57 10.21 &  --33 28 35.7 & 0.68 & --3.60 $\pm$ 0.20 & 16 & 30 & 45 & 44 & 33 & 52 & 50\\
\hline 
ID comp. star & $\alpha$ & $\delta$ &  & [Fe/H]$_{\textnormal{\scriptsize{HR}}}$ & \multicolumn{7}{c}{S/N in wavelength region around (\AA)}  \\
   & (J2000)  &  (J2000) &  & (dex)          & 4000 & 4600 & 5000 & 5300 & 6000 & 7000 & 8700\\
\hline
CS 22891-209 & 19 42 02.16 & --61 03 44.6 & -- & --3.29 $\pm$ 0.14 & 95 & 165 & 250 & 265 & 220 & 200 & 280 \\
CS 22897-008 & 21 03 11.85 & --65 05 08.8 & -- & --3.41 $\pm$ 0.15 & 70 & 155 & 145 & 160 & 125 & 150 & 170 \\
CS 29516-024 & 22 26 15.35 & +02 51 46.2  & -- & --3.06 $\pm$ 0.10 & 50 & 100 & 115 & 135 & 110 & 130 & 150 \\
\hline
\end{tabular}
\end{table*}

\subsection{From equivalent width to abundance}
We derive abundances from the measured equivalent widths of the lines using \textit{Turbospectrum} \citep{alva98}, which is updated consistently with the recently revised version of the (OS)MARCS atmosphere models \citep[e.g.,][]{gust08,plez08} and includes a full treatment of scattering in the source function. The (OS)MARCS atmosphere models we use have a 1D spherical symmetry. The model linelist is created using the Vienna Atomic Line Database (VALD) for the atomic species \citep[e.g.,][]{kupk00} and additionally the contributions from CN, NH, OH and CH molecular lines are modeled \citep[][B. Plez private communication]{spit05}. For the given stellar parameters, \textit{Turbospectrum} calculates the abundance of the element to create a line of the measured equivalent width and provides an error on that abundance using the input error in equivalent width. Synthetic spectra created with \textit{Turbospectrum} are also used to derive abundances through a comparison by eye for the stronger lines mentioned above and the CH band. The solar abundance mixture used is taken from \citet{grev98}. 

\subsection{Final abundances and their uncertainties}
IRAF \textit{splot} provides the option to estimate errors in the equivalent widths of the lines by means of a Monte Carlo simulation. For each spectrum we estimated the signal-to-noise (S/N) at several wavelengths (see Table \ref{tab:targets}), using \textit{splot} to compute the RMS over a continuum region. These S/N values are subsequently used as input to the Gaussian line measuring routine, which creates one thousand simulations around the modeled line given the noise level. The corresponding error, $\Delta$EW as given in Table \ref{tab:linelist_targets} in the online material, is the absolute one sigma (68.3\%) deviation of the parameter estimates. 

\begin{table*}
\caption{Derived abundances for all spectra.\label{tab:abundances}}
\begin{tabular}{cccccccc|cccccccc}
\hline
\hline
X & $N_{X}$ & $\langle$[X/H]$\rangle$ & ab(X) & $\langle$[X/Fe]$\rangle$ & $\sigma_{\textnormal{\scriptsize{EW}}}$ & $\sigma_{X}$ & $\sigma_{fin}$ & X & $N_{X}$ & $\langle$[X/H]$\rangle$ & ab(X) & $\langle$[X/Fe]$\rangle$ & $\sigma_{\textnormal{\scriptsize{EW}}}$ & $\sigma_{X}$ & $\sigma_{fin}$  \\
\hline
\hline
\multicolumn{8}{c|}{Scl002\_06} & \multicolumn{8}{c}{Scl074\_02}\\
\hline
Na  & 2 & --3.68 & 2.65 & --0.32 & 0.20 & 0.05 & 0.20 & Na  & 2 & --3.15 & 3.18 & --0.15 & 0.18 & 0.06 & 0.28\\
Mg  & 4 & --3.15 & 4.43 & 0.20  & 0.12 & 0.18 & 0.13 & Mg  & 4 & --2.83 & 4.75 &  0.17 & 0.15 & 0.09 & 0.20\\
Ca  & 3 & --3.06 & 3.30 & 0.30  & 0.11 & 0.09 & 0.14 & Ca  & 5 & --2.74 & 3.63 &  0.26 & 0.20 & 0.16 & 0.20\\
Ti1 & 0 & ---    & --   & --    & --   & --   & --   & Ti1 & 0 & --    & --   & --    & --   & --   & --  \\
Ti2 & 5 & --3.18 & 1.84 & 0.17  & 0.15 & 0.22 & 0.15 & Ti2 & 2 & --3.00 & 2.02 & 0.00 & 0.46 & 0.15 & 0.46\\
Cr  & 2 & --3.95 & 1.72 & --0.59 & 0.10 & 0.06 & 0.18 & Cr  & 1 & --3.59 & 2.08 & --0.59 & 0.11 & --   & 0.39 \\
Fe1 & 28& --3.36 & 4.15 &  0.00 & 0.11 & 0.25 & 0.05 & Fe1 & 25& --3.00 & 4.50 &  0.00 & 0.18 & 0.39 & 0.08\\
Fe2 & 3 & --3.08 & 4.42 &  0.27 & 0.19 & 0.19 & 0.19 & Fe2 & 2 & --3.23 & 4.27 & --0.23 & 0.21 & 0.03 & 0.28\\
Ni  & 1 & --3.61 & 2.65 & --0.24 & 0.09 & --   & 0.25 & Ni  & 0 & --    & --   &   --  & --   & --   & --  \\
Sr  & 1 & --4.22 &--1.25 & --0.86 & 0.50 & --   & 0.50 & Sr  & 1 & --3.84 &--0.87 & --0.84 & 0.30 & --   & 0.39 \\
Ba  & 1 & $<$--4.80 &$<$--2.67 &$<$ --1.44 & -- & -- & -- & Ba  & 0 & $<$--4.13& $<$--2.00& $<$--1.13& --    & --   & -- \\
\hline
\multicolumn{8}{c|}{Scl\_03\_170} & \multicolumn{8}{c}{Scl\_25\_031}\\
\hline
Na  & 1 & --2.43 & 3.90 &  0.57 & 0.58 & --   & 0.49   & Na & 1 & --2.88 & 3.45 & 0.29 & 0.29 & --   & 0.39 \\
Mg  & 4 & --2.78 & 4.80 &  0.22 & 0.14 & 0.13 & 0.25 & Mg & 3 & --2.61 & 4.98 & 0.56 & 0.17 & 0.10 & 0.22\\
Ca  & 4 & --2.99 & 3.37 & --0.02 & 0.24 & 0.16 & 0.25 & Ca & 2 & --2.63 & 3.73 & 0.53 & 0.23 & 0.10 & 0.28\\
Ti1 & 1 & --2.36 & 2.66 &  0.64 & 0.18 & --   & 0.49   & Ti1& 2 & --2.51 & 2.51 & 0.66 & 0.22 & 0.15 & 0.28\\
Ti2 & 5 & --2.72 & 2.30 &  0.28 & 0.22 & 0.22 & 0.22 & Ti2& 4 & --2.47 & 2.55 & 0.70 & 0.46 & 0.10 & 0.46\\
Cr  & 0 & --    & --   & --    & --   & --   & --   & Cr & 0 & --    & --   & --   & --   & --   & -- \\
Fe1 & 25& --3.00 & 4.50 &  0.00 & 0.23 & 0.49 & 0.10 & Fe1& 25& --3.17 & 4.33 & 0.00 & 0.26 & 0.39 & 0.08\\
Fe2 & 2 & --2.84 & 4.66 &  0.16 & 0.19 & 0.39 & 0.35   & Fe2 & 2 & --2.59 & 4.91 & 0.58 & 0.37 & 0.40 & 0.37  \\
Ni  & 0 & --    & --   & --   & --    & --   & --   & Ni & 0 & --    & --   & --   & --   & --   & -- \\
Sr  & 2 & --3.02 &--0.05 & -0.02 & 0.30 & --   & 0.35   & Sr & 2 & --2.77 & 0.20 & 0.40 & 0.50 & 0.20 & 0.50\\
Ba  & 2 & --3.63 &--1.50 & --0.63 & 0.20 & 0.02 & 0.35 & Ba & 0 & --    & --   & --    & --   & --   & -- \\
\hline
\multicolumn{8}{c|}{Scl051\_05} & \multicolumn{8}{c}{Scl024\_01}\\
\hline
Na & 1 & --2.93 & 3.40 & --0.47 & 0.20 & --   & 0.40 & Na & 2 & --2.68 & 3.65 & 0.23 & 0.40 & 0.05 & 0.40\\
Mg & 2 & --2.61 & 4.97 & --0.15 & 0.12 & 0.07 & 0.28 & Mg & 4 & --2.46 & 5.12 & 0.44 & 0.23 & 0.13 & 0.23\\
Ca & 3 & --2.70 & 3.67 & --0.24 & 0.26 & 0.23 & 0.26 & Ca & 3 & --2.73 & 3.63 & 0.18 & 0.17 & 0.08 & 0.18\\
Ti1& 0 & --    & --   & --   & --    & --   & --   & Ti1& 5 & --2.87 & 2.15 & 0.03 & 0.20 & 0.22 & 0.20\\
Ti2& 5 & --2.98 & 2.04 & --0.53 & 0.24 & 0.18 & 0.24 & Ti2 & 5 & --2.51 & 2.51 & 0.40 & 0.25 & 0.16 & 0.25\\
Cr & 2 & --2.64 & 3.03 & --0.18 & 0.30 & 0.30 & 0.30 & Cr & 2 & --3.18 & 2.49 & --0.27 & 0.27 & 0.28 & 0.27\\
Fe1 &28& --2.46 & 5.04 &  0.00 & 0.27 & 0.40 & 0.08 & Fe1&32 & --2.91 & 4.59 & 0.00 & 0.23 & 0.31 & 0.05\\
Fe2& 4 & --2.51 & 4.99 &  -0.05 & 0.24 & 0.34 & 0.24 & Fe2& 3 & --2.70 & 4.80 & 0.21 & 0.22 & 0.23 & 0.22\\
Ni & 0 & --    & --   & --    & --   & --   & --   & Ni & 1 & --3.08 & 3.17 &--0.17 & 0.17 & --   & 0.31  \\
Sr & 1 & --4.67 &--1.70 & --2.21 & 1.30 & --  & 1.30    & Sr & 2 & --2.37 & 0.60 & 0.54 & 0.50 & 0.20 & 0.50 \\
Ba & 0 & $<$--3.83& $<$--1.70& $<$--1.37& --    & --   & --  & Ba & 1 & --3.10 & --0.97 & --0.19 & 0.36   & --   & 0.36  \\
\hline
\multicolumn{8}{c|}{Scl031\_11} & \multicolumn{8}{c|}{}\\
\hline
Na  & 1 & --4.13 & 2.20 & --0.66 & 0.15 & --  & 0.29 & & & & & & & & \\
Mg  & 3 & --3.66 & 3.92 & --0.19 & 0.19 & 0.13 & 0.19 & & & & & & & & \\
Ca  & 3 & --3.58 & 2.78 & --0.11 & 0.86 & 0.18 & 0.86 & & & & & & & & \\
Ti1 & 1 & --3.17 & 1.85 &  0.30 & 0.59 & --   & 0.59   & & & & & & & & \\
Ti2 & 4 & --3.13 & 1.89 &  0.34 & 0.34 & 0.47 & 0.34 & & & & & & & & \\
Cr  & 0 & --    & --   & --    & --   & --   & -- & & & & & & & & \\
Fe1 &18 & --3.47 & 4.03 &  0.00 & 0.29 & 0.29 & 0.07 & & & & & & & & \\
Fe2 & 3 & --3.38 & 4.12 &  0.09 & 0.20 & 0.21 & 0.20 & & & & & & & & \\
Ni  & 1 & --3.91 & 2.34 & --0.44 & 0.45 & --   & 0.45 & & & & & & & & \\
Sr  & 1 & $<$--4.47 & $<$--1.50 & $<$--1.00 & -- & -- & -- & & & & & & & & \\
Ba  & 0 & $<$--3.83 & $<$--1.70 & $<$--0.36 & -- & --   & --    & & & & & & & & \\
\hline
\end{tabular}
\end{table*}

\begin{table*}
\centering
\caption{Changes in the mean value of [X/H] due to uncertainties in the stellar parameters, either derived from an increase in temperature by $100$ K and consistent changes in log(g) and $v_{t}$ ($\Delta$(T,g,v)), or by an increase of log(g) alone of 0.2 dex ($\Delta$g). These errors behave symmetrical by approximation such that a decrease instead of increase of temperature and/or log(g) by the same amount would give similar absolute changes.}
\label{tab:stelerr}
\centering
\begin{tabular}{l|cc|cc|cc|cc|cc|cc|cc}
\hline
\hline
  & \multicolumn{2}{c|}{Scl002\_06} & \multicolumn{2}{c|}{Scl074\_02} & \multicolumn{2}{c|}{Scl\_03\_170} & \multicolumn{2}{c|}{Scl\_25\_031} & \multicolumn{2}{c|}{Scl051\_05} & \multicolumn{2}{c|}{Scl024\_01} & \multicolumn{2}{c}{Scl031\_11}\\
El. & $\Delta$(T,g,v) & $\Delta$g & $\Delta$(T,g,v) & $\Delta$g & $\Delta$(T,g,v) & $\Delta$g & $\Delta$(T,g,v) & $\Delta$g & $\Delta$(T,g,v) & $\Delta$g & $\Delta$(T,g,v) & $\Delta$g & $\Delta$(T,g,v) & $\Delta$g \\
 & (dex) & (dex)&  (dex) & (dex)&  (dex) & (dex)& (dex) & (dex) & (dex) & (dex) & (dex) & (dex) & (dex) & (dex)\\
\hline 
Na & 0.23 & --0.01 & 0.19 & --0.02 & 0.11 & --0.03 & 0.16 & --0.11  & 0.16 & --0.02 & 0.21 & --0.04 & 0.12 & --0.01\\
Mg & 0.16 & --0.03 & 0.14 & --0.04 & 0.09 & --0.04 & 0.12 & --0.06  & 0.15 & --0.07 & 0.15 & --0.08 & 0.11 & --0.02\\
Ca & 0.12 & --0.02 & 0.12 & --0.03 & 0.09 & --0.03 & 0.09 & --0.01  & 0.09 & --0.02 & 0.17 & --0.06 & 0.11 & --0.03\\
Ti1 & -- & -- & -- & --            & 0.14 & --0.01 & 0.14 & --0.01  & --   & --     & 0.16 & --0.01 & 0.15 & --0.01\\
Ti2 & 0.06 & 0.05 & 0.09 & 0.06    & 0.06 & 0.06   & 0.05 & 0.06    & 0.09 & 0.07   & 0.10 & 0.04   & 0.09 & 0.06 \\
Cr & 0.21 & --0.01 & 0.16 & --0.02 & --   & --0.01 & --   & --      & 0.15 & --0.01 & 0.22 & --0.03 & --   & --   \\
Fe1 & 0.26 & 0.01 & 0.25 & --0.01  & 0.13 & --0.01 & 0.12 & --0.002 & 0.22 & --0.03 & 0.25 & --0.01 & 0.15 & --0.01\\
Fe2 & 0.02 & 0.05 & 0.05 & 0.06    & 0.03 & 0.07   & 0.02 & 0.07    & 0.06 & 0.07   & 0.04 & 0.06   & 0.05 & 0.06 \\
Ba & 0.11 & 0.07 & -- & --         & 0.10 & 0.06   & --   & --  &     --   & --     & 0.24 & 0.02   & --   & --   \\
\hline
\end{tabular}
\end{table*}

The S/N changes considerably over the whole wavelength range, generally being lower at shorter wavelengths. In our calculation of final abundances and errors for elements for which more than one line could be measured, all measurements are weighed by 1/$\Delta$EW$^{2}$. All weights are normalized so that their sum (per element in each star) equals 1.0. The final abundance is taken as the weighted mean of all measurements for each element. All these values can be found in Table \ref{tab:abundances}, along with three different error estimates. The first is the weighted average error in equivalent width, $\sigma_{\textnormal{\scriptsize{EW}}}$, the second the weighted dispersion in the derived abundances of a given element, $\sigma_{x}$. For [\ion{Fe}{I}/H] for which a significant number of lines are present, we see that $\sigma_{x}$ is larger than $\sigma_{\textnormal{\scriptsize{EW}}}$. This is unsurprising since the latter just takes into account the error on the equivalent width parameter, while the former also reflects other uncertainties such as the continuum level. From our analysis of the halo stars in common with \citet{cayr04}, we find that the dispersion in the \ion{Fe}{I} measurements with excellent S/N ratios is $\sim$0.17 dex. As expected, the dispersion in the Sculptor stars is somewhat higher than this value and becomes even higher for the stars with lower S/N. The final error for [\ion{Fe}{I}/H] is calculated as $\sigma_{FeI}$/$\sqrt{N_{FeI}}$. However, for elements with just a few measurements, such an error may underestimate the true error due to low-number statistics. Therefore we take the dispersion of Fe measurements, $\sigma_{Fe}$, as a minimum and we take $\sigma_{\textnormal{\scriptsize{EW}}}$ also as a minimum value for the error, such that the final error is described by: max($\sigma_{\textnormal{\scriptsize{EW}}}$, max($\sigma_{\textnormal{\scriptsize{FeI}}}$,$\sigma_{x}$)/$\sqrt{N_{x}}$).    

The final error does not include effects due to the uncertainties in the derived stellar parameters from photometry, as such systematic errors are hard to estimate. In Table \ref{tab:stelerr} we illustrate the effect of changes in effective temperature, surface gravity and microturbulence on the derived abundances of [X/H]. Since both the derived gravity and microturbulence depend on temperature, we derive one error from an increase of the temperature by 100K and consistent changes in gravity and microturbulence ($\Delta$(T,g,v)). We note here that comparison between either several temperature scales or different colors in Section \ref{sec:stelparam} typically resulted in differences $<$ 100K. Additionally, we derive an error from an increase of the log(g) by 0.2 dex ($\Delta$g) to incorporate uncertainties in the derived surface gravity due to, for instance, the unknown exact stellar mass of each star. Generally for [\ion{Fe}{I}/H] the stellar parameter changes produced by a shift of 100K dominate over the final error in the element measurement (compare Table \ref{tab:stelerr} to Table \ref{tab:abundances}) because of the many \ion{Fe}{I} lines available. However, for a few stars both errors are comparable in magnitude. For abundances of most other elements and [X/Fe], the measurement error dominates.

\subsection{Non-LTE effects}
All our abundances are derived using the assumption of Local Thermodynamic Equilibrium (LTE). However, for many elements calculations of non-LTE effects have been carried out in the literature \citep[e.g.,][]{mash07,andr07,mash08,andr10,berg10,andr11,lind11,mash11,berg12,lind12}. For most elements the non-LTE corrections are expected to be small compared to the error bars given in our study. These corrections are dependent on many parameters (temperature, gravity, strength of the line) implying that non-LTE abundances should be modeled individually for each star and line, which is beyond the scope of this work. An exception are the non-LTE corrections for Na, which have only a negligible dependence on temperature and gravity of the star in the low-metallicity regime \citep{andr07}. We therefore apply non-LTE corrections to our Na abundances in Section \ref{sec:na}.

\section{Abundances for the comparison halo stars}\label{sec:cayrab}

\begin{figure}
\includegraphics[width=\linewidth]{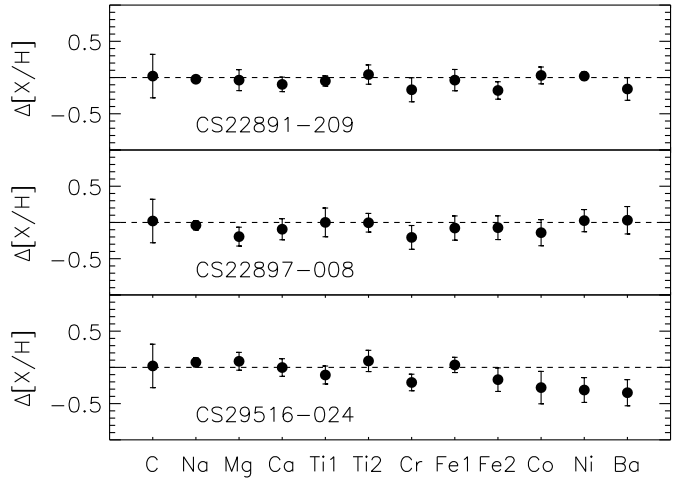}
\caption{The difference in [X/H] between our analysis from X-shooter spectra and that by \citet{cayr04}, \citet{fran07} and \citet{boni09} on high-resolution, high signal-to-noise spectra of the same stars. All elements considered are listed on the x-axis. \label{fig:cayruscomp}}  
\end{figure}

In Figure \ref{fig:cayruscomp} we show a comparison of our abundances from the X-shooter spectra for the three halo stars taken from the sample of \citet{cayr04}.  The results from the high-resolution analysis in \citet{cayr04} are supplemented with \citet{fran07} for the heavy elements and \citet{boni09} for the updated Mg abundances. Overall there is good agreement which lends confidence to both our linelist and the method of deriving abundances. We have used the temperatures, gravities and microturbulences as given in \citet{cayr04} for the derivation of abundances. Since \citet{cayr04} also use \textit{Turbospectrum} and (OS)MARCS models for their analysis, no uncertainties in the stellar parameters or modeling are tested here. However, the difference in resolving power between our spectra is a factor $\sim$8 and our spectra additionally have a lower S/N. Using synthetic modeling of the CH band around 4300 \AA, we also derive the same best fit values for [C/Fe].

\section{Abundances for the Sculptor targets}\label{sec:sclab}

\begin{figure}
\includegraphics[width=\linewidth]{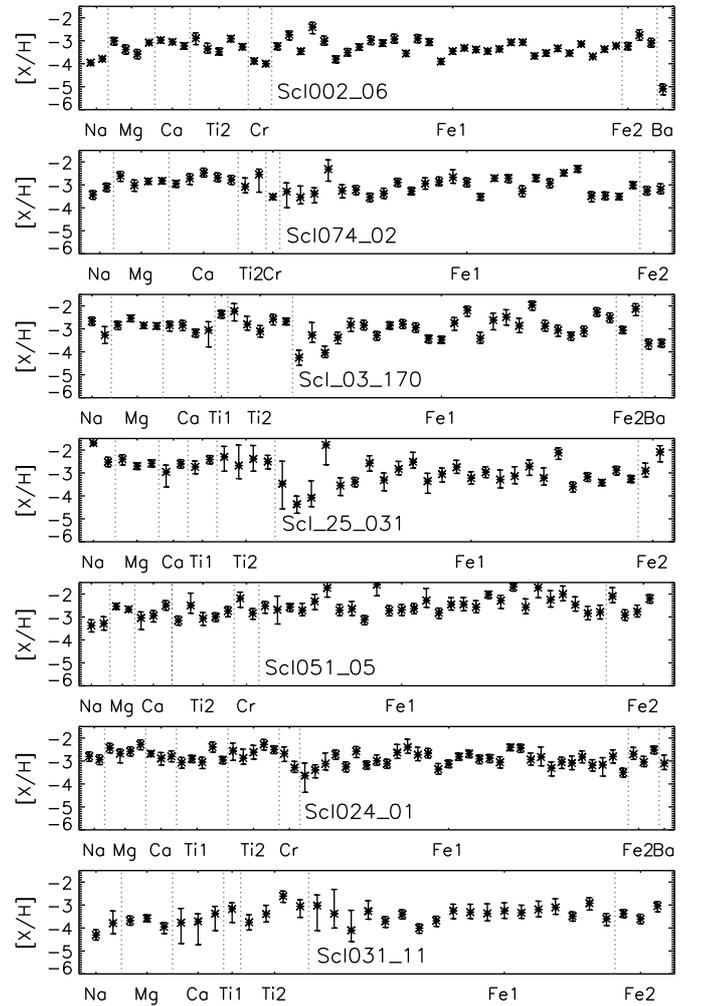}
\caption{Abundances for all Sculptor stars sorted per element, and by wavelength within each element, ranging from blue to red. \label{fig:abundances}}
\end{figure}

Figure \ref{fig:abundances} shows the abundances for all Sculptor targets on a line-by-line basis, sorted per element and on central wavelength within each element. Such a representation is very helpful to detect possible problems in the reduction phase, which might show up as slopes in the determination for abundances in the blue and red parts of the spectrum, or undetected blends which will make one line to stand out in each of the spectra. None of these spurious features can be found in Figure  \ref{fig:abundances}, the derived abundances are clearly well-behaved. The effect of lower S/N in the blue end of the spectrum is clearly present, resulting in higher dispersion for each element.

\subsection{Metallicity}

\subsubsection{Iron}
A dominant element in stellar spectra by its abundance of absorption lines, iron is synthesized by many different nucleosynthetic processes and produced and released in supernovae explosions. Type-Ia SNe contribute especially large relative amounts of iron and iron-peak elements. 

In our spectra we determine \ion{Fe}{I} abundances from 18 to 32 lines in each spectrum, most of which are in the UVB. Although all are very metal-poor, five stars have [Fe/H]$=-$3 or below and are thus confirmed to be extremely metal-poor. Interestingly, the five candidates shown in Figure \ref{fig:CMDtargetsel} that lie somewhat offset from the locus of the RGB (where we would expect the most extremely metal-poor giant branch to lie, but also the lower asymptotic giant branch), have the lowest [FeI/H] within our sample. The lowest [FeI/H] measured is [FeI/H]=-3.47 $\pm$ 0.07 for Scl031\_11. 

The abundances derived from \ion{Fe}{I} and \ion{Fe}{II} lines are in good agreement with each other, taking into account the error bars and small number of \ion{Fe}{II} lines (2-4 lines for each star). The good agreement between these two ionization stages of Fe also indicates that our values for the surface gravities, which were derived from photometry in Section \ref{sec:stelparam}, are appropriate. As illustrated in Table \ref{tab:stelerr}, the effect of a 0.2 dex variation in the surface gravities on our abundance results is very small (on the order of $\sim$0.05 dex). Since our derived error bars on the \ion{Fe}{II} abundances are many times larger than this, we conclude that the derivation of spectroscopic surface gravities, through balancing \ion{Fe}{I} and \ion{Fe}{II} abundances, would be of limited value. Additionally, the balance between \ion{Fe}{I} and \ion{Fe}{II} abundances is known to be affected by non-LTE effects which will vary on a star-by-star and line-by-line basis. 

In Scl00\_06 and Scl\_25\_031 the difference between [\ion{Fe}{II}/H] and [\ion{Fe}{I}/H] is somewhat larger than 1$\sigma$ (although $<$1.5$\sigma$). As these stars show [\ion{Fe}{II}/H]$>$[\ion{Fe}{I}/H] this might be (partly) explained by overionization from non-LTE effects, which is known to be able to cause differences \citep{mash11,berg12,lind12}. Based on the non-LTE correction dependence on temperature, surface gravity and metallicity shown in \citet{lind12}, we estimate that we might expect a non-LTE correction for our stars between 0.07-0.15 dex for high-excitation unsaturated \ion{Fe}{I} lines. This is the right order of magnitude to explain any differences which cannot be solely explained by the size of the error bars. As [FeI/H] is derived from a much larger collection of lines than [FeII/H], we use this as our default iron measurement and will refer to [FeI/H] as [Fe/H] for the remainder of the paper. 

\subsubsection{Comparison to the CaT predictions}

\begin{figure}
\includegraphics[width=\linewidth]{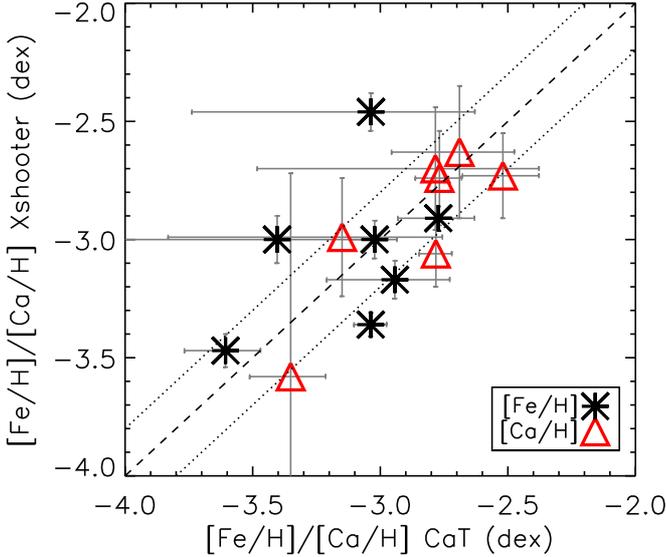}
\caption{CaT predictions for either [Fe/H] and [Ca/H] from the calibrations of \citet{star10} versus the abundances for those elements derived directly from the element absorption lines in the X-shooter spectra.\label{fig:CaT}}
\end{figure}

Figure \ref{fig:CaT} summarizes the agreement of the abundances derived in this study compared with the predictions from the CaT lines. This work presents the most extensive test to date of this new calibration in the extremely metal-poor regime, therefore it is important to look at it in more detail. 

Both the calibration from CaT strength to [Fe/H] and [Ca/H] are used \citep[see the appendix of][]{star10} and compared to direct measurements of [Fe/H] and [Ca/H] from our X-shooter spectra. The calibration to Ca shows less dispersion, suggesting that the CaT lines do trace calcium better than iron in these particular stars. This is consistent with the behavior we find in the models \citep{star10}. However, we know from other abundance studies that for higher metallicities, for instance within the Fornax dwarf galaxy, a better agreement with [Fe/H] is found \citep{batt08a}. Since all metals can influence the equivalent width of the CaT lines through their contribution to the electron pressure in the star \citep{shet09}, it is expected that the CaT lines do not strictly follow provided calibration relations given in \citet{star10} in all cases. The calibrations are derived using fixed abundance ratios \citep[see][for details]{star10} and stars with deviating chemical abundance patterns would fall off the relation \citep[][for instance show the effect of unusually large Mg variations within NGC 2419 on the CaT line strengths]{mucc12}. With that proviso, our work here clearly shows that the CaT-derived [Fe/H] can be used over a large range of metallicities down to the extremely metal-poor regime, and with an accuracy dependent on the detailed chemical composition of the stars.

\subsection{Carbon}

Carbon can be produced by bringing together $\alpha$-particles in the later stages of nuclear fusion. It is also used as a catalyst in the H-burning phase of the star. In the process of turning H into He the abundances of C and (to a much lesser extend) O are decreased while N is produced. The results of the CNO-cycle only become apparent in the stellar spectrum when the elements from the interior are brought to the surface by mixing processes starting on the RGB \citep[see also][for a direct measure of this depletion in globular clusters]{mart08,shet10}. 

We synthesize the CH molecular band region around 4300 \AA \ and compare this to our spectra in order to measure the [C/Fe] ratios. In the synthetic spectra the same CH linelist as in \citet{cayr04} is used \citep[see also][]{spit05,spit06}. Due to the locking of C in CO or CN molecules, the derived carbon abundances are also sensitive to the oxygen and nitrogen abundance. Since we do not directly measure oxygen or nitrogen abundances for our stars, we use solar [N/Fe] and our derived Mg abundances as an educated guess for the amount of oxygen present. Moderate variations in [N/H] or [O/H], e.g. around 0.4 dex, do not lead to a measurable change in the derived [C/H] with our method. Similarly, reasonable changes in the isotope fractions for carbon do not affect our results significantly. The results shown use a $^{12}$C to $^{13}$C ratio of six, typical for a star on the RGB.

\begin{figure}
\begin{center}
\includegraphics[width=\linewidth]{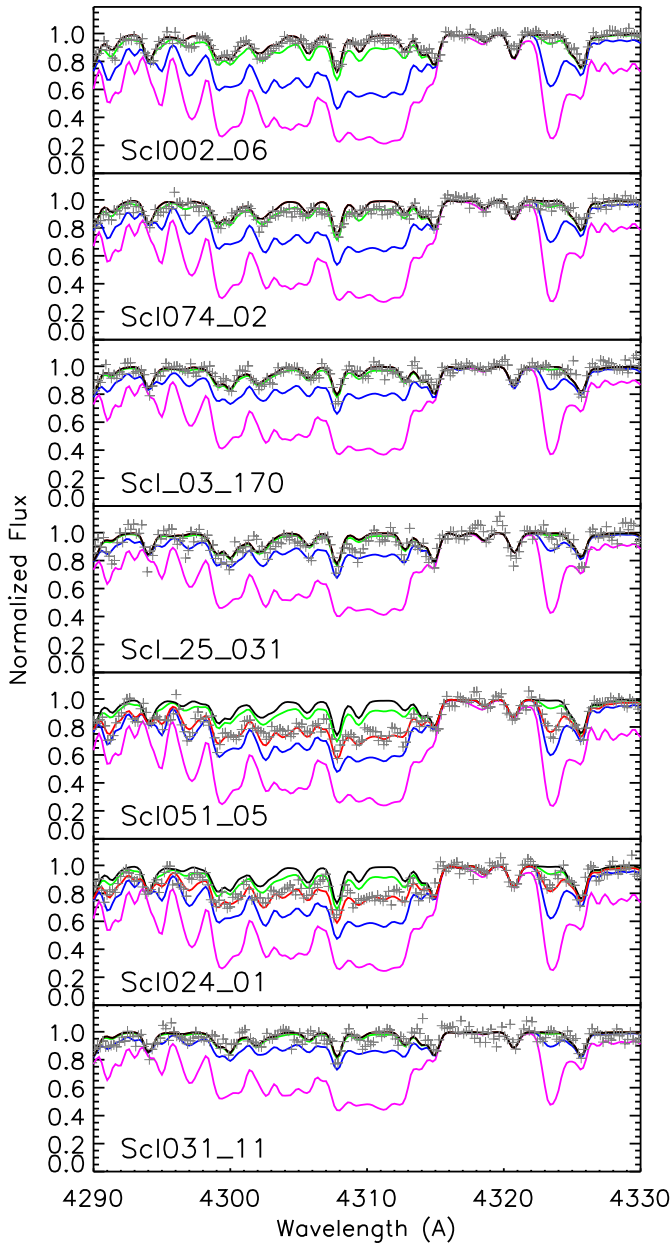}
\caption{The region of the molecular CH band. The gray plusses are the observed spectra for the Sculptor target stars, the colored lines are synthetic spectra for [C/Fe]=$-2.0$ (black), [C/Fe]=$-1.0$ (green), [C/Fe]=$+0.0$ (blue), [C/Fe]=$+1.0$ (magenta). For Scl051\_05 and Scl024\_01, the best fitting synthetic spectra are overplotted in red and have [C/Fe]=$-0.4$ and [C/Fe]=$-0.5$ respectively. \label{fig:Carbon}}
\end{center}
\end{figure}

In this region of the spectrum, we re-normalize the spectra locally using a lower order cubic spline for the continuum to prevent the fit from following the very broad CH feature and thereby mimicking a weaker feature. We match the normalized spectrum to synthetic spectra with a range of [C/Fe] values, as shown in Figure \ref{fig:Carbon}. From this analysis, we find that all of the Sculptor targets are carbon-poor. For most targets the synthetic spectra suggest approximately [C/Fe]= --1, below which it is difficult to make a more precise measurement. The fit to Scl002\_06 clearly prefers a [C/Fe]$< -1$. The best fit values for Scl051\_05 and Scl024\_01 are [C/Fe]=--0.4 and [C/Fe]=--0.5 respectively, which makes them the least Carbon depleted stars in the sample. We adopt a minimal error in the derived [C/Fe] values of $\pm$0.3 dex, which is mainly reflecting the continuum placement uncertainty as this can have a significant effect on the value derived.

\begin{figure}
\begin{center}
\includegraphics[width=\linewidth]{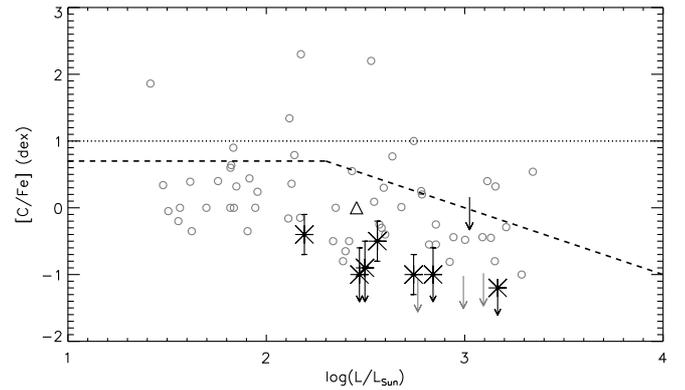}
\caption{[C/Fe] versus the bolometric luminosity of the target stars (black asterisks). Both the classical definition for carbon-rich ([C/Fe]$>$1, horizontal dotted line) and the definition from \citet{aoki07} taking into account mixing on the RGB (dashed line) are indicated. One Sculptor star from \citet{tafe10} is shown as a large black open triangle. The upper limit for the Sculptor star of \citet{freb10a} is shown as a black arrow and upper limits for Carina stars from \citep{venn12} are shown as gray arrows. Stars from other dwarf galaxies are shown as open gray circles \citep{fulb04,cohe09,tafe10,norr10c,norr10a,hond11,lai11}. \label{fig:Carbonvslum}}
\end{center}
\end{figure}

Figure \ref{fig:Carbonvslum} shows the obtained [C/Fe] versus the bolometric luminosity of the star (assuming a stellar mass of 0.8M$_{\odot}$). Clearly, none of our stars are carbon-rich, in the classical sense of having [C/Fe]$>$1, as indicated by the dotted line. \citet{aoki07} argue that mixing of CNO processes and therewith the depletion of carbon, should be taken into account when assessing whether a star is carbon-rich. The dashed line in Figure \ref{fig:Carbonvslum} indicates the dividing line between carbon-rich and carbon-normal metal-poor stars according to their definition. None of our stars match their carbon-rich criterion. 

In the Galactic halo the percentage of extremely metal-poor stars which are carbon-rich is estimated to be $14-32$\% depending on the sample and definition used  \citep{norr97a,ross99,chri03,mars05,cohe05,luca06,yong12b}. Some of these stars show enhancements in s-process elements as well, indicating they might be part of a binary system with a companion star that went through the AGB phase causing pollution with material rich in carbon and s-process elements. However, the existence of a significant number of carbon-rich extremely metal-poor stars with no s-process enhancements or different abundance patterns indicates another mechanism must produce these stars \citep[e.g.][]{norr97b,aoki02,beer05}. 

So far studies of small samples of (extremely) metal-poor stars in dwarf galaxies have only found a few carbon-rich stars \citep{norr10a,hond11,lai11}, as illustrated in Figure \ref{fig:Carbonvslum}. For the classical dwarf galaxies, one very carbon-rich star has been found in the Sextans dwarf galaxy \citep{hond11}. On the other hand the three stars studied in detail in the classical Carina dwarf galaxy are carbon-poor \citep{venn12}. In our work we find no carbon-rich stars in Sculptor, which could indicate that the mechanism responsible for the enhancement in carbon is dependent on environment. 

We further investigate whether this result could still be driven by our limited sample size. Currently there are nine stars with measurements, or upper limits, of [C/Fe] in the extremely metal-poor regime of the Sculptor dwarf galaxy, seven from this work, one from \citet{tafe10} and an upper limit from the star in \citet{freb10a} \citep[for the other star from][no carbon abundance was measured]{tafe10}. None of these are carbon-rich in the sense of showing [C/Fe]$>$1, although the upper limit for the star of \citet{freb10a} is just above the criterion of \citet{aoki07} for carbon-rich stars. The probability of finding no carbon-rich stars when studying nine stars in the extremely metal-poor regime is low if a significant percentage of the population is carbon-rich. We perform a simple likelihood calculation, assuming that 20\% of the extremely metal-poor stars should be carbon-rich and that our set of extremely metal-poor stars represents a 
fair sample from this larger population, a plausible assumption for this relatively large dwarf spheroidal. The probability can in this case simply be written as P$_{0}=(1-r)^{n}$, where $r$ is the carbon-rich fraction and $n$ the number of stars observed. For $r$=0.20, this yields a probability of $\sim$13\% to find no carbon-rich stars in a sample of nine stars. If we assume that instead of 20\% up to 32\% of all extremely metal-poor stars should be carbon-rich, as suggested by the recent work of \citet{yong12b} using the carbon-rich definition of \citet{aoki07}, the probablity of it being a chance effect shrinks to 3\% (5\% when we only use the eight stars with direct carbon measurements). The measurement of \citet{yong12b} was derived from a strictly [Fe/H]$<$-3.0 sample and in general the fraction of carbon-rich stars is thought to decrease with increasing metallicities. However, several studies report very high fractions for samples with [Fe/H]$<-2.5$ or even [Fe/H]$<-2.0$ \citep[e.g.][]{mars05,luca06}, suggesting that the fraction changes only moderately in this regime.

We can also perform a similar calculation using the finite dataset of the Sculptor dwarf galaxy from DART. This dataset of 629 stars contains 24 extremely metal-poor candidates (defined here as having a CaT-based estimate of [Fe/H]$<$-2.75). Of these 24 stars we have followed up eight \citep[the star from][is not included in the DART dataset]{freb10a} for which we were able to study carbon and none of these was found to be carbon-rich. If we assume the actual amount of carbon-rich stars in the DART sample of 24 stars to be 5 (corresponding to $\sim$20\%), the probability of finding none in our sample of eight would have been 10\%. In the case that 8 out of 24 are actually carbon-rich \citep[corresponding to the $\sim$32\% from][]{yong12b} the probability becomes 2\%.

These numbers all hint that the underlying distributions of carbon-rich stars might be different in the Sculptor dwarf galaxy compared to the Milky Way environment, although this requires confirmation using larger samples. 

\subsection{Alpha elements}

\textit{$\alpha$-Elements} are produced from multiples of $\alpha$-particles during the various burning stages in intermediate mass or massive stars and are dispersed in the interstellar medium mainly by SN II explosions. Large samples of stars in dwarf galaxies and the Galactic halo have clearly demonstrated that the dwarfs show a lower [$\alpha$/Fe] at metallicities [Fe/H]$>-2$ dex than the Galactic halo stars at similar metallicity. This is generally explained as being due to the onset of significant Type-Ia supernovae contribution at a different metallicity in various systems, which donate far more iron than $\alpha$-elements to the interstellar medium causing [$\alpha$/Fe] to decrease in next generations of stars \citep[e.g.,][]{venn04,tols09}. 

\begin{figure*}
\includegraphics[width=\linewidth]{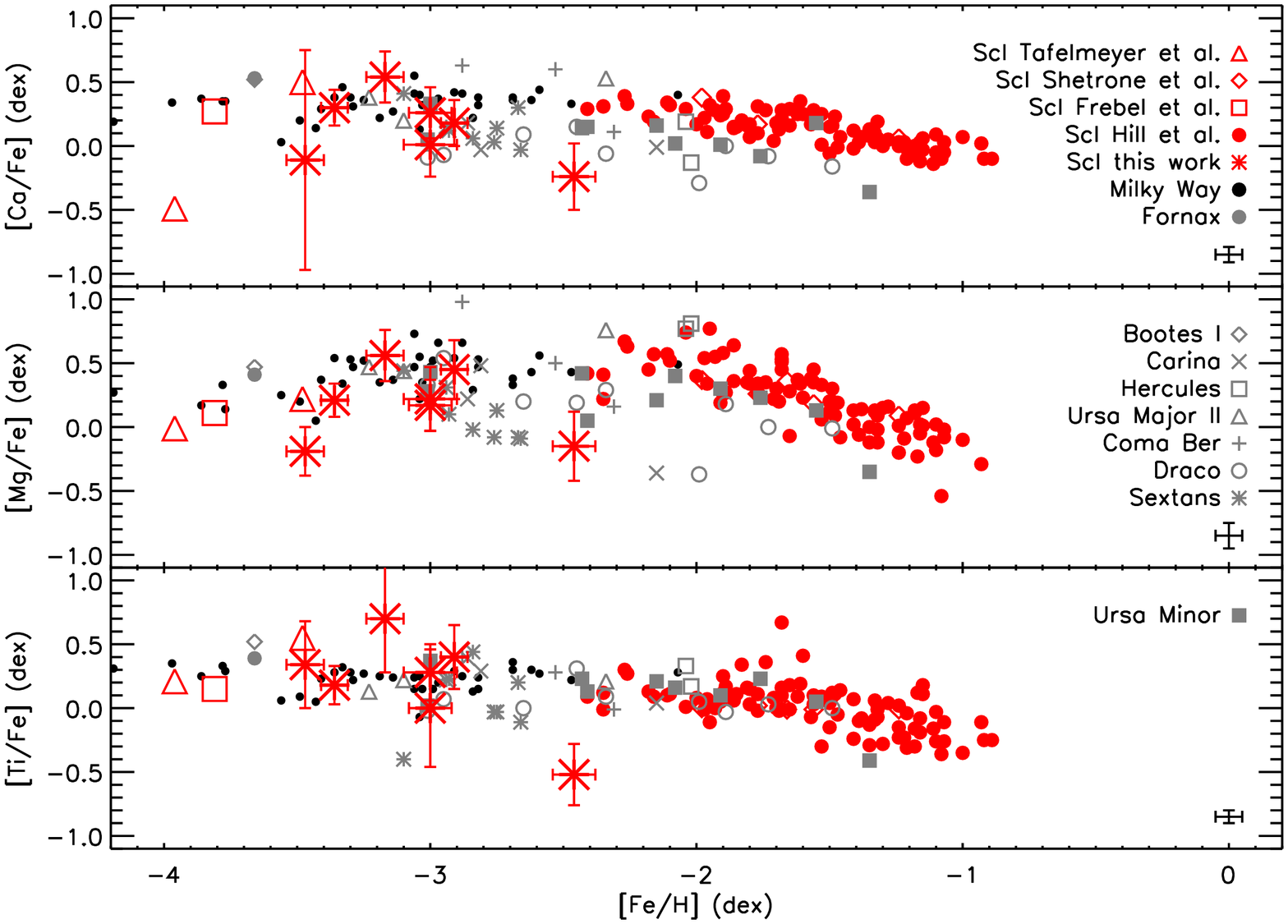}
\caption{[Ca/Fe], [Mg/Fe] and [Ti/Fe] abundances for the Sculptor targets from this work versus their [Fe/H] (red asterisks). Overplotted are stars from other studies. Red symbols indicate Sculptor stars from Hill et al., in preparation, \citet{shet03}, \citet{tafe10} and \citet{freb10a}. The black small circles are the halo stars from the \citet{cayr04} sample. A typical error bar for this sample is shown in the lower right corner of each panel. The Mg values are taken as updated by \citet{boni09}. A sample of very and extremely metal-poor stars from other galaxies are plotted as gray symbols. We show all literature values for stars in ultra-faint galaxies and stars with [Fe/H]$<$--2.5 dex in the classical dwarf spheroidals. \citet{freb10b} for Ursa Major II and Coma Berenices, \citet{tafe10} for Fornax, \citet{tafe10} and \citet{aoki09} for Sextans, \citet{venn12}, and \citet{lema12}, for Carina, \citet{koch08} for Hercules, \citet{cohe09} and \citet{fulb04} for Draco, \citet{cohe10} for Ursa Minor and \citet{norr10b} for Bootes I. For Ti, \ion{Ti}{II} abundances are used for the stars from this work, since these are determined with smaller errors generally. For most other works we adopt \ion{Ti}{II} abundances as well, except for the Hercules stars and the Draco star from \citet{fulb04} where only \ion{Ti}{I} was available. For the \citet{cayr04} sample, the average of \ion{Ti}{I} and \ion{Ti}{II} is shown. All [Fe/H] abundances are determined from \ion{Fe}{I} lines, except again for \citet{cayr04} stars for which we show an average value (generally the difference between their \ion{Fe}{I} and \ion{Fe}{II} abundances is very minimal).\label{fig:alpha}}
\end{figure*}

The main alpha elements which can be observed in our spectra are Mg, Ca and Ti. Their abundances are plotted in Figure \ref{fig:alpha} in comparison with the abundances for the same elements in other samples of (extremely) metal-poor stars in the Milky Way, including other dwarf galaxies and additional stars in the Sculptor dwarf galaxy. Only the work of \citet{cayr04} is shown for Galactic stars, since this is a very homogeneous and large dataset and is almost devoid of carbon-rich stars (as is our sample) which often show chemical peculiarities. Our Sculptor stars show a similar distribution, but arguably a larger spread than the halo stars of similar metallicity. The largest dispersion is seen in [Ti/Fe]. The two most distinct outliers show the same trend in Ca and Mg as in Ti. However, we caution that for most of the elements the spread is consistent within the error bars. 

Scl051\_05 (with [Fe/H]=$-2.46$ dex) has much lower alpha element abundances compared to the halo sample of similar [Fe/H]. This is the faintest star in our sample, but the signal-to-noise reached is comparable with that of the other stars. Compared to the sample of very metal-poor stars from other dwarf galaxies, the discrepancy is somewhat reduced since there are more stars observed with a similar [Fe/H] and lower alpha abundances. In particular, \citet{venn12} report the existence of a very [$\alpha$/Fe] low star in the Carina dwarf spheroidal and conclude from a thorough analysis of many other elements that this particular star is probably iron-rich due to its birth inside a pocket of SNIa enriched gas. This peculiar chemistry pattern has also been seen in several stars in the Galactic halo \citep{ivan03,cayr04,yong12a}. Additionally, \citet{tafe10} point out that in their sample there are signs of inhomogeneous mixing, as two stars of similar metallicity in the Sextans dwarf galaxy show very different [Mg/Fe] and [C/Fe] abundances. From the limited number of elements available for study in this work, Ca, Mg, Ti, Na, Sr and Ba are all underabundant in Scl051\_05 with respect to the Milky Way pattern, the iron-peak element Cr is not. This would support a scenario in which this star is relatively enhanced in SNIa products. 

Our lowest [Fe/H] star has an unusually low Mg abundance compared to all samples, but the sample of stars measured outside the Milky Way halo at this [Fe/H] is particularly small. We do note here that \citet{aoki09} do report a low [Mg/Fe] for several very metal-poor stars in the Sextans dwarf galaxy. In this regard, the unusual low Ca abundance observed in the lowest metallicity Scl star from \citet{tafe10} is intruiging. The authors caution however that for this particular star, the Ca abundance is measured from only one very strong line, which could be affected by non-LTE and is in the low S/N region of the spectrum.

\subsection{Na abundances}{\label{sec:na}}

The non-LTE Na abundances for the Milky Way halo stars follow a flat relation. This is theoretically expected at low metallicities when Na is mainly produced in massive stars during their carbon-burning phase. At higher metallicities the abundance of Na also depends on the amount of neutron-rich elements and becomes metallicity dependent \citep[e.g.][]{woos95}. In higher metallicity samples ([Fe/H]$>-1.5$ dex) of the dwarf spheroidals, [Na/Fe] lies very significantly below [Na/Fe] of the Milky Way components at similar metallicity \citep[see][for an overview]{tols09}. As for the lower abundance of alpha elements, this is explained by the low impact of massive stars on the chemical enrichment of the dSphs.

\begin{figure}[!h]
\begin{center}
\includegraphics[width=\linewidth]{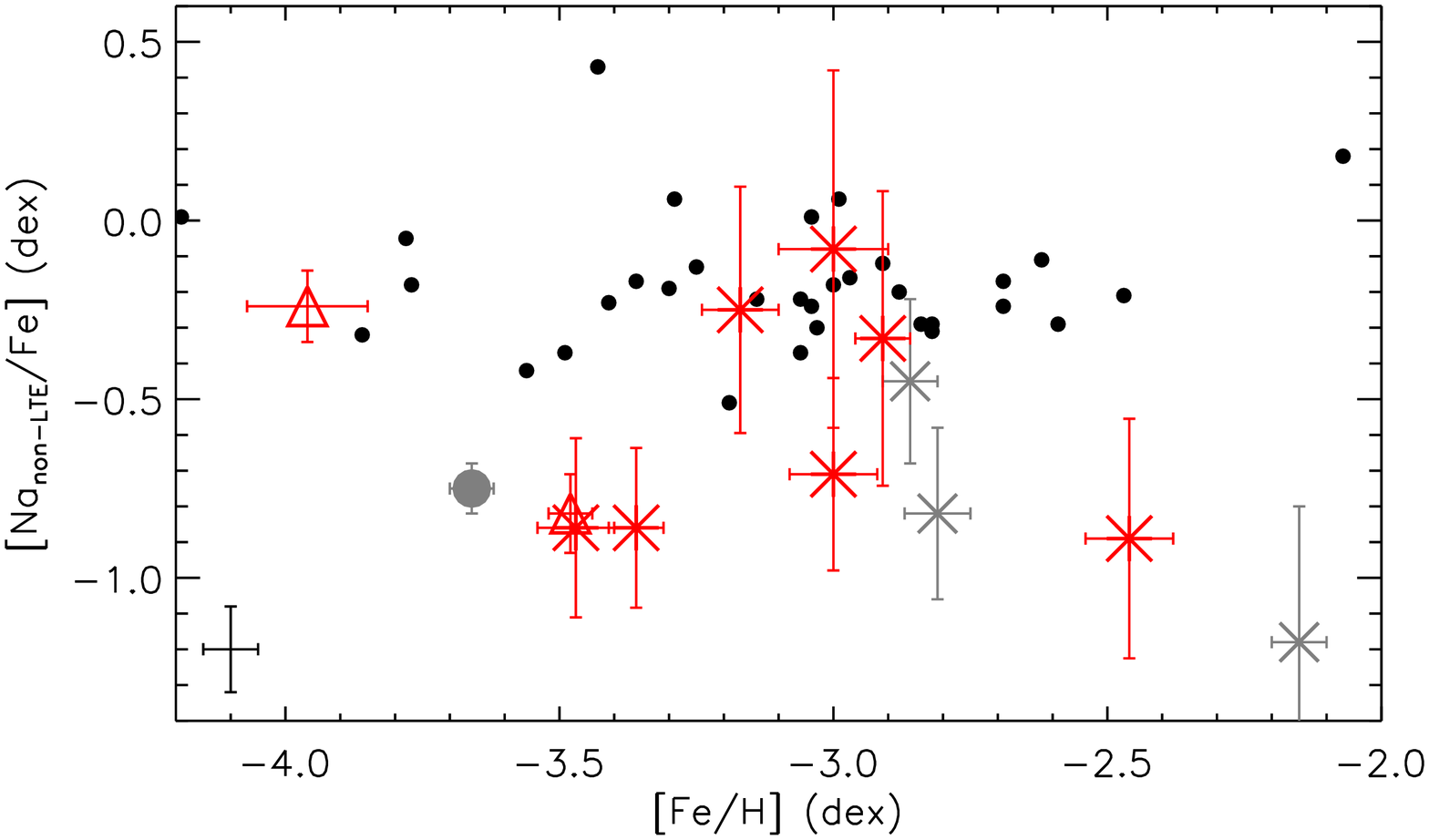}
\caption{[Na/Fe] corrected for non-LTE versus [Fe/H] for our Sculptor sample (red asterisks), extremely metal-poor halo stars \citep[][black filled circles]{andr07}, two other stars in the Sculptor dSph and one in the Fornax dSph from \citet{tafe10} and three stars in Carina from \citet{venn12} (red triangles, gray circle and gray crosses respectively). A typical error bar for the Milky Way stars is shown in the bottom left corner of the figure. \label{fig:nafe}}
\end{center}
\end{figure}

We derive the Na abundances for our stars from the strong resonance Na D1 and D2 lines. For a sample of extremely metal-poor halo stars, \citet{andr07} have published curves of equivalent width versus non-LTE corrections. Our equivalent widths are all within the range of their study allowing us to adopt the \citet{andr07} values to correct our Na abundances for non-LTE effects. The corrections applied are listed in Table \ref{tab:Nanonlte}. For some of our stars the [Fe/H]=$-3$ dex curve did not reach high enough equivalent widths, in which case we used the [Fe/H]=$-2.5$ dex curve. An extra error of 0.1 dex is quadratically added to the LTE Na abundances to account for all the uncertainties in the correction. Figure \ref{fig:nafe} shows the resulting non-LTE abundances together with the halo stars from \citet{andr07}, two stars in Sculptor and one in Fornax from \citet{tafe10} and three stars in Carina from \citet{venn12}, all corrected using the same curves. Most of the stars follow the Milky Way relation, but two of our most metal-poor stars and our least metal-poor star lie significantly below the Milky Way relation, as do several stars from the other works shown. Note that the low [Na/Fe] stars in our sample also show low [Mg/Fe] values.

\begin{table}
\caption{Non-LTE corrections from \citet{andr07} for the Na D1 line (5889.9 \AA) and the Na D2 line (5895.9 \AA).}
\label{tab:Nanonlte}
\centering
\begin{tabular}{l|ccc}
\hline
\hline
 star & corr. D1 & corr. D2 \\
\hline 
 Scl002\_06 & --0.57 & --0.50 \\
 Scl074\_02 &  --0.53 & --0.59 \\
 Scl\_03\_170 & --0.65\tablefootmark{1} & -- \\
 Scl\_25\_031 & -- & --0.54\tablefootmark{1} \\
 Scl051\_05 & --0.42 & --  \\
 Scl024\_01 & --0.50\tablefootmark{1} & --0.61\tablefootmark{1} \\
 Scl031\_11 & --0.20 & -- \\
\hline
\end{tabular}
\tablefoot{\tablefoottext{1}{Corrections were obtained from the relation for models with [Fe/H]=--2.5 dex instead of [Fe/H]=--3.0 dex.}}
\end{table}

\begin{figure}[!h]
\begin{center}
\includegraphics[width=\linewidth]{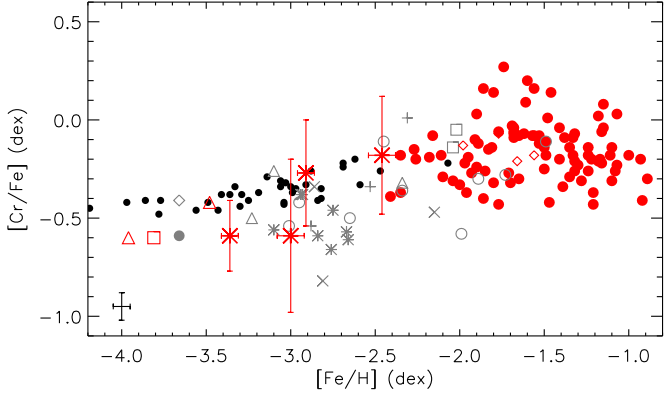}
\caption{[Cr/Fe] versus [Fe/H] for our Sculptor program stars and results from other works for Sculptor stars and very and extremely low-metallicity stars in other galaxies. References, color coding and symbols are the same as in Figure \ref{fig:alpha}. A typical error bar for the Milky Way stars is shown in the bottom left corner of the figure. \label{fig:crfe}}
\end{center}
\end{figure}

\subsection{Iron peak elements: Cr}
Of the iron-peak elements, Cr is the only one which could be measured reliably in our spectra besides Fe. One Co line is also present, but since this line is in the blue part of the spectrum and within the wings of the very broad CH-band, the placement of the continuum is difficult, making the abundance derivation from this one weak line very uncertain. One Ni line could be measured reliably only in three stars. Chromium is known to follow a very narrow trend with [Fe/H], almost flat in the very and extremely low-metallicity regime. Our results follow this trend within error bars, as is shown in Figure \ref{fig:crfe}.

\subsection{Heavy elements: Ba and Sr}

All elements heavier than Zn can no longer be produced by nuclear fusion and are instead created by neutron capture processes. These can take place in two modes: \textit{slow} when exposed to relatively low neutron densities so that the timescale for capturing a neutron is larger than that of a typical decay time, or \textit{rapid} at high neutron densities. Hence the two modes are called s- and r-process. The s-process takes place in low and intermediate-mass stars at the end of their lives and in the helium and carbon burning phases of more massive stars. Where exactly the r-process takes place is not well understood. Most heavy elements can be created through both channels since they have various stable isotopes that can be reached by either one of the processes.

\begin{figure}
\begin{center}
\includegraphics[width=\linewidth]{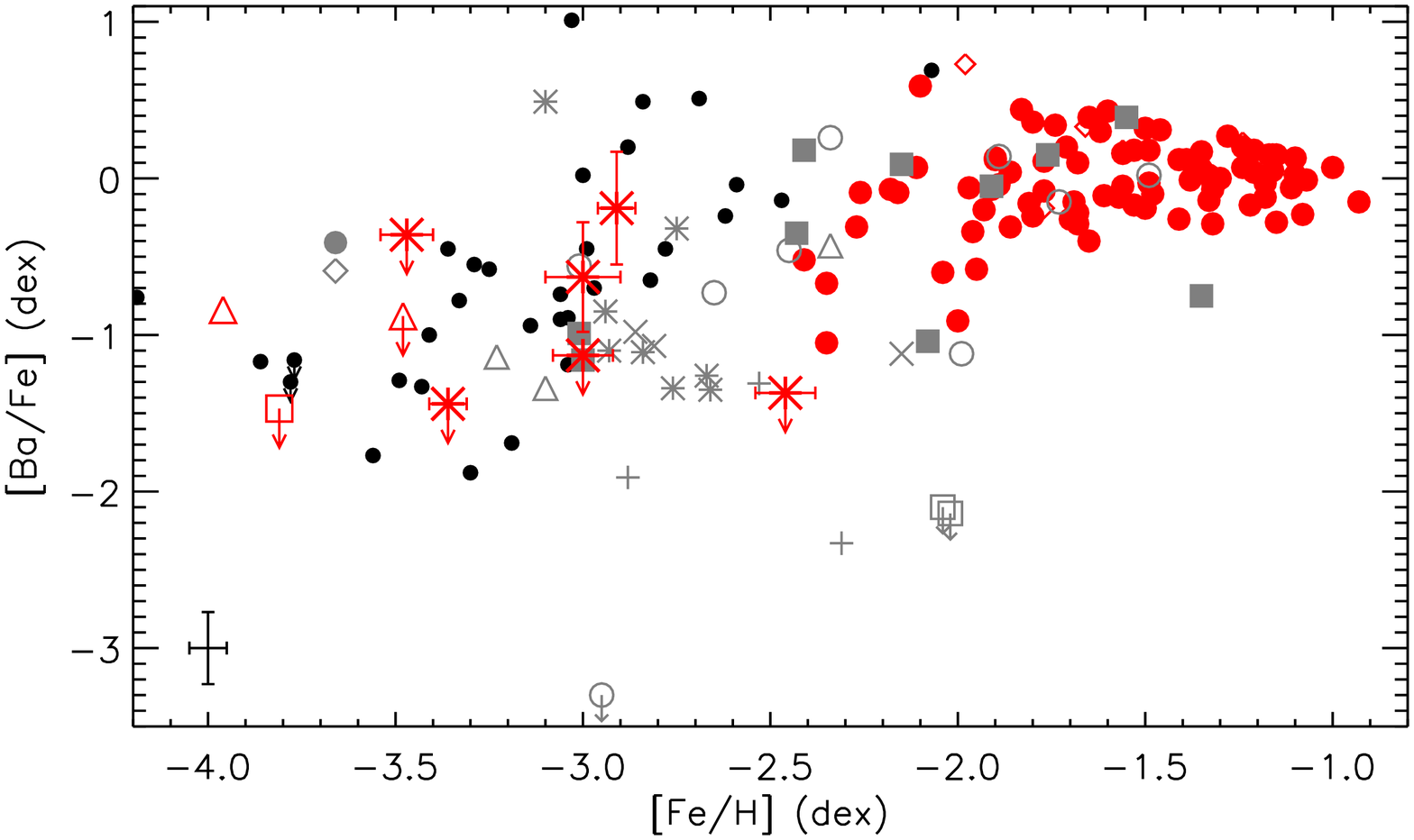}
\caption{[Ba/Fe] versus [Fe/H] for our Sculptor program stars and results from other works for Sculptor stars and very and extremely low-metallicity stars in other galaxies. References, color coding and symbols are the same as in Figure \ref{fig:alpha}, except for the Milky Way sample for which Ba abundances are measured by \citet{fran07}. All symbols which coincide with an error pointing down are derived upper limits. A typical error bar for the Milky Way stars is shown in the bottom left corner of the figure. \label{fig:ba}}
\end{center}
\end{figure}

\begin{figure}
\begin{center}
\includegraphics[width=\linewidth]{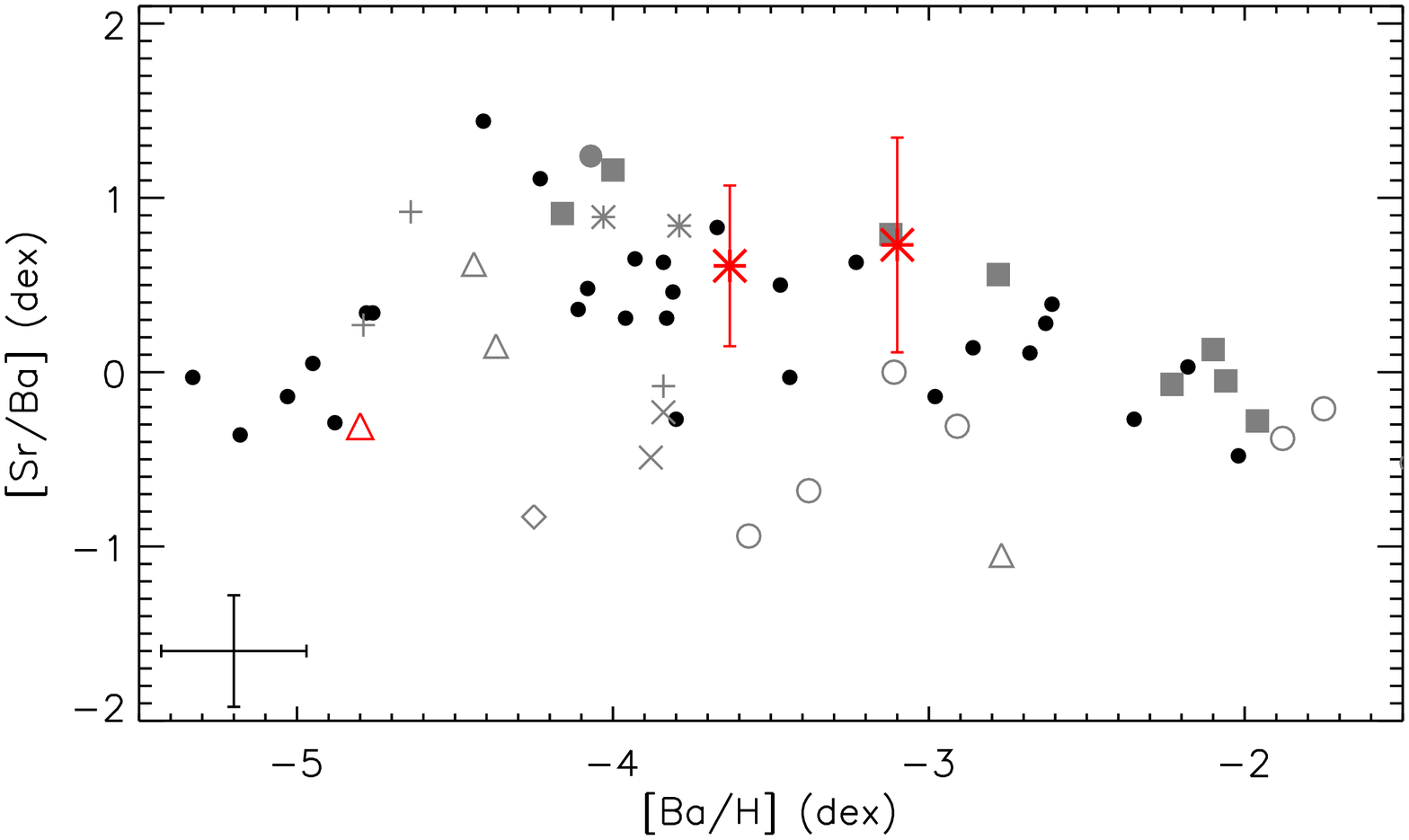}
\caption{[Sr/Ba] versus [Ba/H] for our Sculptor program stars and results from other works for Sculptor stars and very and extremely low-metallicity stars in other galaxies. References, color coding and symbols are the same as in Figure \ref{fig:alpha}, except for the Milky Way sample for which Ba and Sr abundances are measured by \citet{fran07}. A typical error bar for the Milky Way stars is shown in the bottom left corner of the figure. \label{fig:srba}}
\end{center}
\end{figure}

Ba and Sr, the two heavy elements we can measure in our moderate resolution and modest S/N spectra, are both considered to be dominated by s-process creation at solar abundances. In the metal-poor regime, however, both originate from r-process production rather than s-process. In Figure \ref{fig:ba} we show that [Ba/Fe] is low for all our stars, but not unusually low as compared to the other samples shown. Generally we see similar behavior in the full Sculptor sample to that in the Milky Way halo -- little scatter for [Fe/H]$>$--2, but increasing significantly towards lower metallicities. Comparing at --2.7$<$[Fe/H]$<$--2.2 the high-resolution sample of Hill et al., (in prep., red filled circles in Figure \ref{fig:ba}) to the Milky Way sample, there is a hint that the Sculptor stars are at lower [Ba/Fe] values \citep[see also][]{tols09}. However, larger sample of stars with -3.0 $<$ [Fe/H] $<$ -2.0 in Sculptor are needed to investigate this further against a larger comparison sample of Milky Way stars. The current sparsity of data at these metallicities in the Sculptor galaxy is entirely due to selection effects for the high-resolution samples.   

Although [Ba/H] shows a large dispersion at extremely low metallicities, a much narrower trend can be observed in halo stars if one plots [Sr/Ba] vs. [Ba/H]. This is usually interpreted as evidence for a weak r-process which produces only the lighter heavy elements such as Sr (but not Ba) very early in the Universe, thus explaining the quick rise of [Sr/Ba] which then curves down again as the main r-process takes over \citep{fran07}. In our spectra two strong Sr lines can be observed in the blue wavelength regions. Unfortunately, the abundances derived from these lines have generally larger error bars, since both of the lines are blends. In Figure \ref{fig:srba} we show [Sr/Ba] vs. [Ba/H] for those stars in our sample for which both Sr and Ba could be measured. Within their large error bars, they follow the narrow trend defined by the halo stars. Interestingly, and as has been noticed before \citep{tafe10,venn12}, not all stars in the dwarf galaxies seem to follow the narrow Galactic trend, in particular several ultra-faint galaxies as well as Draco and Carina stars lie significantly below.

\section{Conclusions}\label{sec:Xshootconc}

We have presented direct measurements of Fe, Na, Mg, Ca, Ti, Cr, Sr and Ba abundances in seven extremely metal-poor candidates in the Sculptor dwarf spheroidal which were selected from CaT samples \citep{batt08a}. This work has clearly shown that CaT line data combined with the most recent calibration \citep{star10} can be used as a metallicity indicator down to the extremely metal-poor regime. All extremely metal-poor candidates are confirmed to have [Fe/H]$<-2.5$ dex from direct measurements of Fe lines. Five stars are in the extremely metal-poor regime with [Fe/H]$<-3.0$ dex. Our lowest metallicity star has [Fe/H]=--3.5. 

This is the largest sample to date of extremely metal-poor stars studied in a galaxy external to the Milky Way. It is insightful to compare the abundance patterns to those of Milky Way halo stars, as any differences could provide clues on the history of the galaxy and early star formation in different environments. In general, the trends follow the patterns observed in the Milky Way much more closely than the higher metallicity stars in the Sculptor dwarf galaxy, particularly when comparing the $\alpha$-element abundances. There is a hint of higher scatter in the abundances in Sculptor, nonetheless it remains unclear if this indicates a more stochastic and less well mixed environment in the Sculptor dwarf spheroidal than in the Milky Way, or simply poorer quality data. 
One star shows significant low [X/Fe] in comparison to the Milky Way pattern for all elements except the iron-peak element Cr. This is consistent with a relatively large contribution from type-Ia SNe products. 

Intrguiging hints for different evolution from the halo population are found in the lighter elements. All our stars prove to be carbon-poor, which is intruiging given that many ($\sim$20\%-32\%) of the halo stars at this metallicity are enriched in carbon. We estimate there is a low probability ($\sim$2--13\%) that this is entirely a chance effect. Larger samples are clearly needed to shed light on the formation of carbon-rich stars and their dependence on environment. Additionally, several stars in our sample are more than 1$\sigma$ away from the trend of [Na/Fe] found in the Milky Way halo. 

The earliest stages of the evolution of heavy elements in Sculptor look very similar to the halo pattern. A more complete sample around [Fe/H]=--2.5 will be needed to follow the precise trend of [Ba/Fe] within the Sculptor dSph. The two Sculptor giants for which we could measure both Sr and Ba do follow the less dispersed relation of Milky Way halo stars for [Sr/Ba] vs. [Ba/H]. Also [Cr/Fe] matches the narrow relation of the Milky Way halo stars. 

The data supports a scenario in which the chemical pattern of $\alpha$- and heavy elements in the very first stages of star formation are less dependent on the imprint of the environment than later stages. Whether the differences we find in the lighter elements, Na and C, and the inhomogeneity in the $\alpha$-elements are significant and related to the dwarf spheroidal environment remains to be investigated in larger samples and/or other dwarf spheroidal galaxies. 

\begin{acknowledgements}
We would like to thank Bertrand Plez for making his linelists and Turbospectrum code available and Amina Helmi, Matthew Shetrone and Hugues Sana for very helpful suggestions and discussions. E.S., E.T., L.B., and T.d.B. gratefully acknowledge Netherlands Foundation for Scientific Research (NWO) and the Netherlands Research School for Astronomy (NOVA) for financial support. We thank the International Space Science Institute (ISSI) at Bern for their funding of the team ``Defining the full life-cycle of dwarf galaxy evolution: the Local Universe as a template''. E.S. also gratefully acknowledges the Canadian Institute for Advanced Research (CIfAR) Junior Academy and a Canadian Institute for Theoretical Astrophysics (CITA) National Fellowship for partial support.  V.H. acknowledges the financial support of the Programme National Galaxies (PNG) of the Institut National des Sciences de l'Univers (INSU). The research leading to these results has received funding from the European Union Seventh Framework Program (FP7/2007-2013) under grant agreement number PIEF-GA-2010-274151. We thank the referee for providing useful suggestions to improve the paper.
\end{acknowledgements}

\bibliographystyle{aa}
\bibliography{ms}

\begin{thebibliography}{91}
\expandafter\ifx\csname natexlab\endcsname\relax\def\natexlab#1{#1}\fi

\bibitem[{{Alonso} {et~al.}(1999){Alonso}, {Arribas}, \&
  {Mart{\'{\i}}nez-Roger}}]{alon99}
{Alonso}, A., {Arribas}, S., \& {Mart{\'{\i}}nez-Roger}, C. 1999, \aaps, 140,
  261

\bibitem[{{Alvarez} \& {Plez}(1998)}]{alva98}
{Alvarez}, R. \& {Plez}, B. 1998, \aap, 330, 1109

\bibitem[{{Andrievsky} {et~al.}(2011){Andrievsky}, {Spite}, {Korotin}, {Fran{\c
  c}ois}, {Spite}, {Bonifacio}, {Cayrel}, \& {Hill}}]{andr11}
{Andrievsky}, S.~M., {Spite}, F., {Korotin}, S.~A., {et~al.} 2011, \aap, 530,
  A105

\bibitem[{{Andrievsky} {et~al.}(2010){Andrievsky}, {Spite}, {Korotin}, {Spite},
  {Bonifacio}, {Cayrel}, {Fran{\c c}ois}, \& {Hill}}]{andr10}
{Andrievsky}, S.~M., {Spite}, M., {Korotin}, S.~A., {et~al.} 2010, \aap, 509,
  A88

\bibitem[{{Andrievsky} {et~al.}(2007){Andrievsky}, {Spite}, {Korotin}, {Spite},
  {Bonifacio}, {Cayrel}, {Hill}, \& {Fran{\c c}ois}}]{andr07}
{Andrievsky}, S.~M., {Spite}, M., {Korotin}, S.~A., {et~al.} 2007, \aap, 464,
  1081

\bibitem[{{Aoki} {et~al.}(2009){Aoki}, {Arimoto}, {Sadakane}, {Tolstoy},
  {Battaglia}, {Jablonka}, {Shetrone}, {Letarte}, {Irwin}, {Hill}, {Francois},
  {Venn}, {Primas}, {Helmi}, {Kaufer}, {Tafelmeyer}, {Szeifert}, \&
  {Babusiaux}}]{aoki09}
{Aoki}, W., {Arimoto}, N., {Sadakane}, K., {et~al.} 2009, \aap, 502, 569

\bibitem[{{Aoki} {et~al.}(2007){Aoki}, {Beers}, {Christlieb}, {Norris}, {Ryan},
  \& {Tsangarides}}]{aoki07}
{Aoki}, W., {Beers}, T.~C., {Christlieb}, N., {et~al.} 2007, \apj, 655, 492

\bibitem[{{Aoki} {et~al.}(2002){Aoki}, {Norris}, {Ryan}, {Beers}, \&
  {Ando}}]{aoki02}
{Aoki}, W., {Norris}, J.~E., {Ryan}, S.~G., {Beers}, T.~C., \& {Ando}, H. 2002,
  \apj, 567, 1166

\bibitem[{{Barklem} {et~al.}(2005){Barklem}, {Christlieb}, {Beers}, {Hill},
  {Bessell}, {Holmberg}, {Marsteller}, {Rossi}, {Zickgraf}, \&
  {Reimers}}]{bark05}
{Barklem}, P.~S., {Christlieb}, N., {Beers}, T.~C., {et~al.} 2005, \aap, 439,
  129

\bibitem[{{Battaglia} {et~al.}(2008b){Battaglia}, {Helmi}, {Tolstoy}, {Irwin},
  {Hill}, \& {Jablonka}}]{batt08b}
{Battaglia}, G., {Helmi}, A., {Tolstoy}, E., {et~al.} 2008b, \apjl, 681, L13

\bibitem[{{Battaglia} {et~al.}(2008a){Battaglia}, {Irwin}, {Tolstoy}, {Hill},
  {Helmi}, {Letarte}, \& {Jablonka}}]{batt08a}
{Battaglia}, G., {Irwin}, M., {Tolstoy}, E., {et~al.} 2008a, \mnras, 383, 183

\bibitem[{{Battaglia} \& {Starkenburg}(2012)}]{batt12}
{Battaglia}, G. \& {Starkenburg}, E. 2012, \aap, 539, A123

\bibitem[{{Beers} \& {Christlieb}(2005)}]{beer05}
{Beers}, T.~C. \& {Christlieb}, N. 2005, \araa, 43, 531

\bibitem[{{Bergemann} \& {Cescutti}(2010)}]{berg10}
{Bergemann}, M. \& {Cescutti}, G. 2010, \aap, 522, A9

\bibitem[{{Bergemann} {et~al.}(2012){Bergemann}, {Lind}, {Collet}, {Magic}, \&
  {Asplund}}]{berg12}
{Bergemann}, M., {Lind}, K., {Collet}, R., {Magic}, Z., \& {Asplund}, M. 2012,
  \mnras, 427, 27

\bibitem[{{Bessell} \& {Norris}(1984)}]{bess84}
{Bessell}, M.~S. \& {Norris}, J. 1984, \apj, 285, 622

\bibitem[{{Bonifacio} {et~al.}(2009){Bonifacio}, {Spite}, {Cayrel}, {Hill},
  {Spite}, {Fran{\c c}ois}, {Plez}, {Ludwig}, {Caffau}, {Molaro}, {Depagne},
  {Andersen}, {Barbuy}, {Beers}, {Nordstr{\"o}m}, \& {Primas}}]{boni09}
{Bonifacio}, P., {Spite}, M., {Cayrel}, R., {et~al.} 2009, \aap, 501, 519

\bibitem[{{Cayrel} {et~al.}(2004){Cayrel}, {Depagne}, {Spite}, {Hill}, {Spite},
  {Fran{\c c}ois}, {Plez}, {Beers}, {Primas}, {Andersen}, {Barbuy},
  {Bonifacio}, {Molaro}, \& {Nordstr{\"o}m}}]{cayr04}
{Cayrel}, R., {Depagne}, E., {Spite}, M., {et~al.} 2004, \aap, 416, 1117

\bibitem[{{Cayrel de Strobel} \& {Spite}(1988)}]{cayr88}
{Cayrel de Strobel}, G. \& {Spite}, M., eds. 1988, IAU Symposium, Vol. 132,
  {The impact of very high S/N spectroscopy on stellar physics: proceedings of
  the 132nd Symposium of the International Astronomical Union held in Paris,
  France, June 29-July 3, 1987.}

\bibitem[{{Christlieb}(2003)}]{chri03}
{Christlieb}, N. 2003, in Reviews in Modern Astronomy, Vol.~16, Reviews in
  Modern Astronomy, ed. {R.~E.~Schielicke}, 191

\bibitem[{{Cohen} {et~al.}(2008){Cohen}, {Christlieb}, {McWilliam}, {Shectman},
  {Thompson}, {Melendez}, {Wisotzki}, \& {Reimers}}]{cohe08}
{Cohen}, J.~G., {Christlieb}, N., {McWilliam}, A., {et~al.} 2008, \apj, 672,
  320

\bibitem[{{Cohen} \& {Huang}(2009)}]{cohe09}
{Cohen}, J.~G. \& {Huang}, W. 2009, \apj, 701, 1053

\bibitem[{{Cohen} \& {Huang}(2010)}]{cohe10}
{Cohen}, J.~G. \& {Huang}, W. 2010, \apj, 719, 931

\bibitem[{{Cohen} {et~al.}(2006){Cohen}, {McWilliam}, {Shectman}, {Thompson},
  {Christlieb}, {Melendez}, {Ramirez}, {Swensson}, \& {Zickgraf}}]{cohe06}
{Cohen}, J.~G., {McWilliam}, A., {Shectman}, S., {et~al.} 2006, \aj, 132, 137

\bibitem[{{Cohen} {et~al.}(2005){Cohen}, {Shectman}, {Thompson}, {McWilliam},
  {Christlieb}, {Melendez}, {Zickgraf}, {Ram{\'{\i}}rez}, \&
  {Swenson}}]{cohe05}
{Cohen}, J.~G., {Shectman}, S., {Thompson}, I., {et~al.} 2005, \apjl, 633, L109

\bibitem[{{de Boer} {et~al.}(2012){de Boer}, {Tolstoy}, {Hill}, {Saha},
  {Olsen}, {Starkenburg}, {Lemasle}, {Irwin}, \& {Battaglia}}]{debo12}
{de Boer}, T.~J.~L., {Tolstoy}, E., {Hill}, V., {et~al.} 2012, \aap, 539, A103

\bibitem[{{de Boer} {et~al.}(2011){de Boer}, {Tolstoy}, {Saha}, {Olsen},
  {Irwin}, {Battaglia}, {Hill}, {Shetrone}, {Fiorentino}, \& {Cole}}]{debo11}
{de Boer}, T.~J.~L., {Tolstoy}, E., {Saha}, A., {et~al.} 2011, \aap, 528, A119

\bibitem[{{Fran{\c c}ois} {et~al.}(2007){Fran{\c c}ois}, {Depagne}, {Hill},
  {Spite}, {Spite}, {Plez}, {Beers}, {Andersen}, {James}, {Barbuy}, {Cayrel},
  {Bonifacio}, {Molaro}, {Nordstr{\"o}m}, \& {Primas}}]{fran07}
{Fran{\c c}ois}, P., {Depagne}, E., {Hill}, V., {et~al.} 2007, \aap, 476, 935

\bibitem[{{Fran{\c c}ois} {et~al.}(2012){Fran{\c c}ois}, {Monaco}, {Villanova},
  {Catelan}, {Bonifacio}, {Bellazzini}, {Moni Bidin}, {Marconi}, {Geisler}, \&
  {Sbordone}}]{fran12}
{Fran{\c c}ois}, P., {Monaco}, L., {Villanova}, S., {et~al.} 2012, ArXiv
  e-prints 1202.2899

\bibitem[{{Frebel} {et~al.}(2005){Frebel}, {Aoki}, {Christlieb}, {Ando},
  {Asplund}, {Barklem}, {Beers}, {Eriksson}, {Fechner}, {Fujimoto}, {Honda},
  {Kajino}, {Minezaki}, {Nomoto}, {Norris}, {Ryan}, {Takada-Hidai},
  {Tsangarides}, \& {Yoshii}}]{freb05}
{Frebel}, A., {Aoki}, W., {Christlieb}, N., {et~al.} 2005, \nat, 434, 871

\bibitem[{{Frebel} {et~al.}(2010a){Frebel}, {Kirby}, \& {Simon}}]{freb10a}
{Frebel}, A., {Kirby}, E.~N., \& {Simon}, J.~D. 2010a, \nat, 464, 72

\bibitem[{{Frebel} {et~al.}(2010b){Frebel}, {Simon}, {Geha}, \&
  {Willman}}]{freb10b}
{Frebel}, A., {Simon}, J.~D., {Geha}, M., \& {Willman}, B. 2010b, \apj, 708,
  560

\bibitem[{{Fulbright} {et~al.}(2004){Fulbright}, {Rich}, \& {Castro}}]{fulb04}
{Fulbright}, J.~P., {Rich}, R.~M., \& {Castro}, S. 2004, \apj, 612, 447

\bibitem[{{Geha} {et~al.}(2009){Geha}, {Willman}, {Simon}, {Strigari}, {Kirby},
  {Law}, \& {Strader}}]{geha09}
{Geha}, M., {Willman}, B., {Simon}, J.~D., {et~al.} 2009, \apj, 692, 1464

\bibitem[{{Goldoni} {et~al.}(2006){Goldoni}, {Royer}, {Fran{\c c}ois},
  {Horrobin}, {Blanc}, {Vernet}, {Modigliani}, \& {Larsen}}]{goldoni06}
{Goldoni}, P., {Royer}, F., {Fran{\c c}ois}, P., {et~al.} 2006, in Society of
  Photo-Optical Instrumentation Engineers (SPIE) Conference Series, Vol. 6269,
  Society of Photo-Optical Instrumentation Engineers (SPIE) Conference Series

\bibitem[{{Gratton} {et~al.}(2000){Gratton}, {Sneden}, {Carretta}, \&
  {Bragaglia}}]{grat00}
{Gratton}, R.~G., {Sneden}, C., {Carretta}, E., \& {Bragaglia}, A. 2000, \aap,
  354, 169

\bibitem[{{Grevesse} \& {Sauval}(1998)}]{grev98}
{Grevesse}, N. \& {Sauval}, A.~J. 1998, \ssr, 85, 161

\bibitem[{{Gustafsson} {et~al.}(2008){Gustafsson}, {Edvardsson}, {Eriksson},
  {J{\o}rgensen}, {Nordlund}, \& {Plez}}]{gust08}
{Gustafsson}, B., {Edvardsson}, B., {Eriksson}, K., {et~al.} 2008, \aap, 486,
  951

\bibitem[{{Helmi} {et~al.}(2006){Helmi}, {Irwin}, {Tolstoy}, {Battaglia},
  {Hill}, {Jablonka}, {Venn}, {Shetrone}, {Letarte}, {Arimoto}, {Abel},
  {Francois}, {Kaufer}, {Primas}, {Sadakane}, \& {Szeifert}}]{helm06}
{Helmi}, A., {Irwin}, M.~J., {Tolstoy}, E., {et~al.} 2006, \apjl, 651, L121

\bibitem[{{Honda} {et~al.}(2004){Honda}, {Aoki}, {Ando}, {Izumiura}, {Kajino},
  {Kambe}, {Kawanomoto}, {Noguchi}, {Okita}, {Sadakane}, {Sato},
  {Takada-Hidai}, {Takeda}, {Watanabe}, {Beers}, {Norris}, \& {Ryan}}]{hond04}
{Honda}, S., {Aoki}, W., {Ando}, H., {et~al.} 2004, \apjs, 152, 113

\bibitem[{{Honda} {et~al.}(2011){Honda}, {Aoki}, {Arimoto}, \&
  {Sadakane}}]{hond11}
{Honda}, S., {Aoki}, W., {Arimoto}, N., \& {Sadakane}, K. 2011, \pasj, 63, 523

\bibitem[{{Ivans} {et~al.}(2003){Ivans}, {Sneden}, {James}, {Preston},
  {Fulbright}, {H{\"o}flich}, {Carney}, \& {Wheeler}}]{ivan03}
{Ivans}, I.~I., {Sneden}, C., {James}, C.~R., {et~al.} 2003, \apj, 592, 906

\bibitem[{{Kelson}(2003)}]{kelson03}
{Kelson}, D.~D. 2003, \pasp, 115, 688

\bibitem[{{Kirby} \& {Cohen}(2012)}]{kirb12}
{Kirby}, E.~N. \& {Cohen}, J.~G. 2012, ArXiv e-prints

\bibitem[{{Kirby} {et~al.}(2009){Kirby}, {Guhathakurta}, {Bolte}, {Sneden}, \&
  {Geha}}]{kirb09}
{Kirby}, E.~N., {Guhathakurta}, P., {Bolte}, M., {Sneden}, C., \& {Geha}, M.~C.
  2009, \apj, 705, 328

\bibitem[{{Kirby} {et~al.}(2008){Kirby}, {Simon}, {Geha}, {Guhathakurta}, \&
  {Frebel}}]{kirb08}
{Kirby}, E.~N., {Simon}, J.~D., {Geha}, M., {Guhathakurta}, P., \& {Frebel}, A.
  2008, \apjl, 685, L43

\bibitem[{{Koch} {et~al.}(2008){Koch}, {McWilliam}, {Grebel}, {Zucker}, \&
  {Belokurov}}]{koch08}
{Koch}, A., {McWilliam}, A., {Grebel}, E.~K., {Zucker}, D.~B., \& {Belokurov},
  V. 2008, \apjl, 688, L13

\bibitem[{{Kupka} {et~al.}(2000){Kupka}, {Ryabchikova}, {Piskunov}, {Stempels},
  \& {Weiss}}]{kupk00}
{Kupka}, F.~G., {Ryabchikova}, T.~A., {Piskunov}, N.~E., {Stempels}, H.~C., \&
  {Weiss}, W.~W. 2000, Baltic Astronomy, 9, 590

\bibitem[{{Lai} {et~al.}(2008){Lai}, {Bolte}, {Johnson}, {Lucatello}, {Heger},
  \& {Woosley}}]{lai08}
{Lai}, D.~K., {Bolte}, M., {Johnson}, J.~A., {et~al.} 2008, \apj, 681, 1524

\bibitem[{{Lai} {et~al.}(2011){Lai}, {Lee}, {Bolte}, {Lucatello}, {Beers},
  {Johnson}, {Sivarani}, \& {Rockosi}}]{lai11}
{Lai}, D.~K., {Lee}, Y.~S., {Bolte}, M., {et~al.} 2011, \apj, 738, 51

\bibitem[{{Lemasle} {et~al.}(2012){Lemasle}, {Hill}, {Tolstoy}, {Venn},
  {Shetrone}, {Irwin}, {de Boer}, {Starkenburg}, \& {Salvadori}}]{lema12}
{Lemasle}, B., {Hill}, V., {Tolstoy}, E., {et~al.} 2012, \aap, 538, A100

\bibitem[{{Lind} {et~al.}(2011){Lind}, {Asplund}, {Barklem}, \&
  {Belyaev}}]{lind11}
{Lind}, K., {Asplund}, M., {Barklem}, P.~S., \& {Belyaev}, A.~K. 2011, \aap,
  528, A103

\bibitem[{{Lind} {et~al.}(2012){Lind}, {Bergemann}, \& {Asplund}}]{lind12}
{Lind}, K., {Bergemann}, M., \& {Asplund}, M. 2012, \mnras, 427, 50

\bibitem[{{Lucatello} {et~al.}(2006){Lucatello}, {Beers}, {Christlieb},
  {Barklem}, {Rossi}, {Marsteller}, {Sivarani}, \& {Lee}}]{luca06}
{Lucatello}, S., {Beers}, T.~C., {Christlieb}, N., {et~al.} 2006, \apjl, 652,
  L37

\bibitem[{{Marsteller} {et~al.}(2005){Marsteller}, {Beers}, {Rossi},
  {Christlieb}, {Bessell}, \& {Rhee}}]{mars05}
{Marsteller}, B., {Beers}, T.~C., {Rossi}, S., {et~al.} 2005, Nuclear Physics
  A, 758, 312

\bibitem[{{Martell} {et~al.}(2008){Martell}, {Smith}, \& {Briley}}]{mart08}
{Martell}, S.~L., {Smith}, G.~H., \& {Briley}, M.~M. 2008, \aj, 136, 2522

\bibitem[{{Mashonkina} {et~al.}(2011){Mashonkina}, {Gehren}, {Shi}, {Korn}, \&
  {Grupp}}]{mash11}
{Mashonkina}, L., {Gehren}, T., {Shi}, J.-R., {Korn}, A.~J., \& {Grupp}, F.
  2011, \aap, 528, A87

\bibitem[{{Mashonkina} {et~al.}(2007){Mashonkina}, {Korn}, \&
  {Przybilla}}]{mash07}
{Mashonkina}, L., {Korn}, A.~J., \& {Przybilla}, N. 2007, \aap, 461, 261

\bibitem[{{Mashonkina} {et~al.}(2008){Mashonkina}, {Zhao}, {Gehren}, {Aoki},
  {Bergemann}, {Noguchi}, {Shi}, {Takada-Hidai}, \& {Zhang}}]{mash08}
{Mashonkina}, L., {Zhao}, G., {Gehren}, T., {et~al.} 2008, \aap, 478, 529

\bibitem[{{McWilliam}(1998)}]{mcwi98}
{McWilliam}, A. 1998, \aj, 115, 1640

\bibitem[{{Modigliani} {et~al.}(2010){Modigliani}, {Goldoni}, {Royer},
  {Haigron}, {Guglielmi}, {Fran{\c c}ois}, {Horrobin}, {Bristow}, {Vernet},
  {Moehler}, {Kerber}, {Ballester}, {Mason}, \& {Christensen}}]{modi10}
{Modigliani}, A., {Goldoni}, P., {Royer}, F., {et~al.} 2010, in Society of
  Photo-Optical Instrumentation Engineers (SPIE) Conference Series, Vol. 7737,
  Society of Photo-Optical Instrumentation Engineers (SPIE) Conference Series

\bibitem[{{Mucciarelli} {et~al.}(2012){Mucciarelli}, {Bellazzini}, {Ibata},
  {Merle}, {Chapman}, {Dalessandro}, \& {Sollima}}]{mucc12}
{Mucciarelli}, A., {Bellazzini}, M., {Ibata}, R., {et~al.} 2012, ArXiv e-prints
  1208.0195

\bibitem[{{Norris} {et~al.}(2012){Norris}, {Bessell}, {Yong}, {Christlieb},
  {Barklem}, {Asplund}, {Murphy}, {Beers}, {Frebel}, \& {Ryan}}]{norr12}
{Norris}, J.~E., {Bessell}, M.~S., {Yong}, D., {et~al.} 2012, ArXiv e-prints
  1208.2999

\bibitem[{{Norris} {et~al.}(2010a){Norris}, {Gilmore}, {Wyse}, {Yong}, \&
  {Frebel}}]{norr10a}
{Norris}, J.~E., {Gilmore}, G., {Wyse}, R.~F.~G., {Yong}, D., \& {Frebel}, A.
  2010a, \apjl, 722, L104

\bibitem[{{Norris} {et~al.}(1997a){Norris}, {Ryan}, \& {Beers}}]{norr97a}
{Norris}, J.~E., {Ryan}, S.~G., \& {Beers}, T.~C. 1997a, \apj, 488, 350

\bibitem[{{Norris} {et~al.}(1997b){Norris}, {Ryan}, \& {Beers}}]{norr97b}
{Norris}, J.~E., {Ryan}, S.~G., \& {Beers}, T.~C. 1997b, \apjl, 489, L169

\bibitem[{{Norris} {et~al.}(2001){Norris}, {Ryan}, \& {Beers}}]{norr01}
{Norris}, J.~E., {Ryan}, S.~G., \& {Beers}, T.~C. 2001, \apj, 561, 1034

\bibitem[{{Norris} {et~al.}(2010){Norris}, {Wyse}, {Gilmore}, {Yong}, {Frebel},
  {Wilkinson}, {Belokurov}, \& {Zucker}}]{norr10c}
{Norris}, J.~E., {Wyse}, R.~F.~G., {Gilmore}, G., {et~al.} 2010, \apj, 723,
  1632

\bibitem[{{Norris} {et~al.}(2010b){Norris}, {Yong}, {Gilmore}, \&
  {Wyse}}]{norr10b}
{Norris}, J.~E., {Yong}, D., {Gilmore}, G., \& {Wyse}, R.~F.~G. 2010b, \apj,
  711, 350

\bibitem[{{Pietrzy{\'n}ski} {et~al.}(2008){Pietrzy{\'n}ski}, {Gieren},
  {Szewczyk}, {Walker}, {Rizzi}, {Bresolin}, {Kudritzki}, {Nalewajko}, {Storm},
  {Dall'Ora}, \& {Ivanov}}]{piet08}
{Pietrzy{\'n}ski}, G., {Gieren}, W., {Szewczyk}, O., {et~al.} 2008, \aj, 135,
  1993

\bibitem[{{Plez}(2008)}]{plez08}
{Plez}, B. 2008, Physica Scripta Volume T, 133, 014003

\bibitem[{{Ram{\'{\i}}rez} \& {Mel{\'e}ndez}(2005)}]{rami05}
{Ram{\'{\i}}rez}, I. \& {Mel{\'e}ndez}, J. 2005, \apj, 626, 465

\bibitem[{{Rossi} {et~al.}(1999){Rossi}, {Beers}, \& {Sneden}}]{ross99}
{Rossi}, S., {Beers}, T.~C., \& {Sneden}, C. 1999, in Astronomical Society of
  the Pacific Conference Series, Vol. 165, The Third Stromlo Symposium: The
  Galactic Halo, ed. {B.~K.~Gibson, R.~S.~Axelrod, \& M.~E.~Putman}, 264

\bibitem[{{Schlegel} {et~al.}(1998){Schlegel}, {Finkbeiner}, \&
  {Davis}}]{schl98}
{Schlegel}, D.~J., {Finkbeiner}, D.~P., \& {Davis}, M. 1998, \apj, 500, 525

\bibitem[{{Shetrone} {et~al.}(2010){Shetrone}, {Martell}, {Wilkerson}, {Adams},
  {Siegel}, {Smith}, \& {Bond}}]{shet10}
{Shetrone}, M., {Martell}, S.~L., {Wilkerson}, R., {et~al.} 2010, \aj, 140,
  1119

\bibitem[{{Shetrone} {et~al.}(2003){Shetrone}, {Venn}, {Tolstoy}, {Primas},
  {Hill}, \& {Kaufer}}]{shet03}
{Shetrone}, M., {Venn}, K.~A., {Tolstoy}, E., {et~al.} 2003, \aj, 125, 684

\bibitem[{{Shetrone} {et~al.}(2001){Shetrone}, {C{\^o}t{\'e}}, \&
  {Sargent}}]{shet01}
{Shetrone}, M.~D., {C{\^o}t{\'e}}, P., \& {Sargent}, W.~L.~W. 2001, \apj, 548,
  592

\bibitem[{{Shetrone} {et~al.}(2009){Shetrone}, {Siegel}, {Cook}, \&
  {Bosler}}]{shet09}
{Shetrone}, M.~D., {Siegel}, M.~H., {Cook}, D.~O., \& {Bosler}, T. 2009, \aj,
  137, 62

\bibitem[{{Simon} {et~al.}(2011){Simon}, {Geha}, {Minor}, {Martinez}, {Kirby},
  {Bullock}, {Kaplinghat}, {Strigari}, {Willman}, {Choi}, {Tollerud}, \&
  {Wolf}}]{simo11}
{Simon}, J.~D., {Geha}, M., {Minor}, Q.~E., {et~al.} 2011, \apj, 733, 46

\bibitem[{{Spite} {et~al.}(2006){Spite}, {Cayrel}, {Hill}, {Spite}, {Fran{\c
  c}ois}, {Plez}, {Bonifacio}, {Molaro}, {Depagne}, {Andersen}, {Barbuy},
  {Beers}, {Nordstr{\"o}m}, \& {Primas}}]{spit06}
{Spite}, M., {Cayrel}, R., {Hill}, V., {et~al.} 2006, \aap, 455, 291

\bibitem[{{Spite} {et~al.}(2005){Spite}, {Cayrel}, {Plez}, {Hill}, {Spite},
  {Depagne}, {Fran{\c c}ois}, {Bonifacio}, {Barbuy}, {Beers}, {Andersen},
  {Molaro}, {Nordstr{\"o}m}, \& {Primas}}]{spit05}
{Spite}, M., {Cayrel}, R., {Plez}, B., {et~al.} 2005, \aap, 430, 655

\bibitem[{{Starkenburg} {et~al.}(2010){Starkenburg}, {Hill}, {Tolstoy},
  {Gonz{\'a}lez Hern{\'a}ndez}, {Irwin}, {Helmi}, {Battaglia}, {Jablonka},
  {Tafelmeyer}, {Shetrone}, {Venn}, \& {de Boer}}]{star10}
{Starkenburg}, E., {Hill}, V., {Tolstoy}, E., {et~al.} 2010, \aap, 513, A34

\bibitem[{{Tafelmeyer} {et~al.}(2010){Tafelmeyer}, {Jablonka}, {Hill},
  {Shetrone}, {Tolstoy}, {Irwin}, {Battaglia}, {Helmi}, {Starkenburg}, {Venn},
  {Abel}, {Francois}, {Kaufer}, {North}, {Primas}, \& {Szeifert}}]{tafe10}
{Tafelmeyer}, M., {Jablonka}, P., {Hill}, V., {et~al.} 2010, \aap, 524, A58

\bibitem[{{Tolstoy} {et~al.}(2009){Tolstoy}, {Hill}, \& {Tosi}}]{tols09}
{Tolstoy}, E., {Hill}, V., \& {Tosi}, M. 2009, \araa, 47, 371

\bibitem[{{van Dokkum}(2001)}]{vandokkum01}
{van Dokkum}, P.~G. 2001, \pasp, 113, 1420

\bibitem[{{Venn} {et~al.}(2004){Venn}, {Irwin}, {Shetrone}, {Tout}, {Hill}, \&
  {Tolstoy}}]{venn04}
{Venn}, K.~A., {Irwin}, M., {Shetrone}, M.~D., {et~al.} 2004, \aj, 128, 1177

\bibitem[{{Venn} {et~al.}(2012){Venn}, {Shetrone}, {Irwin}, {Hill}, {Jablonka},
  {Tolstoy}, {Lemasle}, {Divell}, {Starkenburg}, {Letarte}, {Baldner},
  {Battaglia}, {Helmi}, {Kaufer}, \& {Primas}}]{venn12}
{Venn}, K.~A., {Shetrone}, M.~D., {Irwin}, M.~J., {et~al.} 2012, \apj, 751, 102

\bibitem[{{Vernet} {et~al.}(2011){Vernet}, {Dekker}, {D'Odorico}, {Kaper},
  {Kjaergaard}, {Hammer}, {Randich}, {Zerbi}, {Groot}, {Hjorth}, {Guinouard},
  {Navarro}, {Adolfse}, {Albers}, {Amans}, {Andersen}, {Andersen}, {Binetruy},
  {Bristow}, {Castillo}, {Chemla}, {Christensen}, {Conconi}, {Conzelmann},
  {Dam}, {de Caprio}, {de Ugarte Postigo}, {Delabre}, {di Marcantonio},
  {Downing}, {Elswijk}, {Finger}, {Fischer}, {Flores}, {Fran{\c c}ois},
  {Goldoni}, {Guglielmi}, {Haigron}, {Hanenburg}, {Hendriks}, {Horrobin},
  {Horville}, {Jessen}, {Kerber}, {Kern}, {Kiekebusch}, {Kleszcz}, {Klougart},
  {Kragt}, {Larsen}, {Lizon}, {Lucuix}, {Mainieri}, {Manuputy}, {Martayan},
  {Mason}, {Mazzoleni}, {Michaelsen}, {Modigliani}, {Moehler}, {M{\o}ller},
  {Norup S{\o}rensen}, {N{\o}rregaard}, {P{\'e}roux}, {Patat}, {Pena}, {Pragt},
  {Reinero}, {Rigal}, {Riva}, {Roelfsema}, {Royer}, {Sacco}, {Santin},
  {Schoenmaker}, {Spano}, {Sweers}, {Ter Horst}, {Tintori}, {Tromp}, {van
  Dael}, {van der Vliet}, {Venema}, {Vidali}, {Vinther}, {Vola}, {Winters},
  {Wistisen}, {Wulterkens}, \& {Zacchei}}]{vern11}
{Vernet}, J., {Dekker}, H., {D'Odorico}, S., {et~al.} 2011, \aap, 536, A105

\bibitem[{{Woosley} \& {Weaver}(1995)}]{woos95}
{Woosley}, S.~E. \& {Weaver}, T.~A. 1995, \apjs, 101, 181

\bibitem[{{Yong} {et~al.}(2012a){Yong}, {Norris}, {Bessell}, {Christlieb},
  {Asplund}, {Beers}, {Barklem}, {Frebel}, \& {Ryan}}]{yong12a}
{Yong}, D., {Norris}, J.~E., {Bessell}, M.~S., {et~al.} 2012a, ArXiv e-prints
  1208.3016

\bibitem[{{Yong} {et~al.}(2012b){Yong}, {Norris}, {Bessell}, {Christlieb},
  {Asplund}, {Beers}, {Barklem}, {Frebel}, \& {Ryan}}]{yong12b}
{Yong}, D., {Norris}, J.~E., {Bessell}, M.~S., {et~al.} 2012b, ArXiv e-prints
  1208.3003

\end{thebibliography}

\end{document}